\documentclass[useAMS,usenatbib]{mnras}

\usepackage{graphicx}
\usepackage[dvipsnames]{xcolor}
\usepackage{physics}
\usepackage{hyperref}
\usepackage{booktabs}
\usepackage{amsmath}
\usepackage{amssymb}
\usepackage{physics}
\usepackage{bm}
\usepackage{acro}
\usepackage{cleveref}
\usepackage{tabularx}
\usepackage{soul} 
\usepackage{subcaption}
\usepackage{siunitx}
\usepackage[T1]{fontenc}

\DeclareRobustCommand{\VAN}[3]{#2}
\let\VANthebibliography\thebibliography
\def\thebibliography{\DeclareRobustCommand{\VAN}[3]{##3}\VANthebibliography}

\newcommand{\HI}{\ensuremath{\mathrm{H}\scriptstyle\mathrm{I}}}
\newcommand{\p}{\ensuremath{\mathrm{\rho}\scriptstyle\mathrm{_{\rm DM}}}}
\newcommand{\df}{\ensuremath{\mathrm{d}\scriptstyle\mathrm{_{fil}}}}

\begin{document}

\title[Galaxy--environment connection from constrained simulations]{The galaxy--environment connection revealed by constrained simulations}
\author[C. Gallagher et al.]{
Catherine Gallagher$^{1}$\thanks{\href{mailto:22c.gallagher@gmail.com}{22c.gallagher@gmail.com}},
Tariq Yasin$^{1}$,
Richard Stiskalek$^{1,2}$,
Harry Desmond$^{3}$,
Matt J. Jarvis$^{1}$
\\
$^1$Astrophysics, University of Oxford, Denys Wilkinson Building, Keble Road, Oxford, OX1 3RH, UK\\
$^{2}$Center for Computational Astrophysics, Flatiron Institute, 162 5th Ave, New York, NY 10010, USA\\
$^{3}$Institute of Cosmology \& Gravitation, University of Portsmouth, Dennis Sciama Building, Portsmouth, PO1 3FX, UK\\
}

\date{Accepted XXX. Received YYY; in original form ZZZ}
\pubyear{2025}

\maketitle

\begin{abstract}
The evolution of galaxies is known to be connected to their position within the large-scale structure and their local environmental density.
We investigate the relative importance of these using the underlying dark matter density field extracted from the \textit{Constrained Simulations in} \texttt{BORG} (\texttt{CSiBORG}) suite of \textit{constrained} cosmological simulations. We define cosmic web environment through both dark matter densities averaged on a scale up to 16 Mpc/$h$, and through cosmic web location identified by applying \texttt{DisPerSE} to the \texttt{CSiBORG} haloes. We correlate these environmental measures with the properties of observed galaxies in large surveys using optical data (from the NASA-Sloan Atlas) and 21-cm radio data (from ALFALFA). We find statistically significant correlations between environment and colour, neutral hydrogen gas (\HI{}) mass fraction, star formation rate and S\'ersic index. Together, these correlations suggest that bluer, star-forming, \HI{} rich, and disk-type galaxies tend to reside in lower density areas, further from filaments, while redder, more elliptical galaxies with lower star formation rates tend to be found in higher density areas, closer to filaments. We find analogous trends with the quenching of galaxies, but notably find that the quenching of low mass galaxies has a greater dependence on environment than the quenching of high mass galaxies. We find that the relationship between galaxy properties and the environmental density is stronger than that with distance to filament, suggesting that environmental density has a greater impact on the properties of galaxies than their location within the larger-scale cosmic web.
\end{abstract}

\begin{keywords}
large-scale structure of Universe -- dark matter -- galaxies: evolution -- galaxies: statistics
\end{keywords}

\section{Introduction}\label{sec:intro}

In the standard model of cosmology, matter evolves from a nearly smooth primordial density field, with small perturbations at the end of the inflationary epoch. These perturbations evolve non-linearly under gravity to form the cosmic web, a tangled structure of nodes, sheets, filaments and voids~\citep{zel1970gravitational}.

Nodes consist of galaxy clusters and superclusters and are the most massive, gravitationally bound structures in the Universe. Voids are vast expanses of extremely low density, containing little gas and very few isolated galaxies. Sheets are membrane-like 2D structures encircling voids. Filaments are long, thread-like structures that form at the intersection of sheets and which connect the nodes. These filaments are made up of galaxies, gas, and dark matter, which can all flow along the filament  towards the nodes. In the last decade, a significant amount of work as been undertaken to understand how galaxy evolution is dependent on the position within this cosmic web \citep[e.g.][]{Tempel_2013,Zhang_2015,Kraljic2021,Tudorache_2022,Barsanti_2023}.

Moreover, such work can help characterise the relationship between luminous galaxies and the underlying dark matter, which is not directly observable. This is vital for obtaining precision constraints on the cosmological model from forthcoming surveys. The statistics of cosmic web structures are also emerging as a potential test of dark energy, which dominates the late-time evolution of large-scale structure~\citep{2017CMPh...2013901N, 2024arXiv240204837B}.

There are a wide variety of proposed mechanisms for how a galaxy's environment may affect properties such as morphology, colour, stellar mass, and metallicity. For example, simulations show haloes of the same mass in different environments have statistically different assembly histories~\citep{Wechselr2002,Maccio2007,Behroozi2019}, which should impact the observed baryonic properties~\citep{2017MNRAS.465.2381M,2018ApJ...853...84Z,2021NatAs...5.1069C}. Gas is expected to flow along cosmic filaments, and so galaxies along filaments may be expected to replenish their gas more readily and thus have higher gas content~\citep{ramsoy2021rivers}. Spheroidal galaxies are also more commonly located in overdense environments such as nodes and along the filaments, which has been linked to both the age of the most massive haloes and the higher likelihood of galaxy mergers in these dense environments ~\citep[e.g.][]{Naab2009,Cappellari2016, Perez2025}. Galaxy mergers may also lead to quenching, which qualitatively matches the observed higher fraction of red galaxies in denser environments~\citep{dressler1980galaxy,Peng2010,McLure2013}. The most massive galaxies in the most massive haloes  are also more likely to host powerful active galactic nuclei (AGN) that may suppress star formation on large scales through their powerful jets~\citep{Bower2006,Croton2006,Schaye2015,Dubois2016,Bower2017}.

    To investigate these effects observationally, large multi-wavelength surveys are required, both to map the large-scale cosmic web itself, and then to also measure properties of the galaxies that reside within this structure. In this work we use the NASA-Sloan Atlas (NSA) survey and the Arecibo Legacy Fast Arecibo L-band Feed Array survey (ALFALFA,~\citealt{haynes2018arecibo}). The NSA is a catalogue of galaxy properties derived from the Sloan Digital Sky Survey (SDSS;~\citealt{2000AJ....120.1579Y, 2009ApJS..182..543A, 2011ApJS..193...29A}), and contains properties including absolute magnitude, star formation rate parameters, metallicity, and stellar mass. ALFALFA is a neutral hydrogen gas (\HI{}) survey. \HI{} is the reservoir of cold gas that condenses to molecular H$_{2}$ to form stars~\citep{palla1983primordial}. Therefore, constraining its relationship with galaxy properties is vital for determining the details of galaxy growth. One complication when studying the link between \HI{} and galaxies is that, because \HI{} resides predominantly in the lowest density outer regions of galaxies, it is highly susceptible to environmental effects. In dense environments, \HI{} is often stripped away, as shown by numerous studies~\citep[e.g.][]{Solanes2001,Cortese2011,Hess2013,Odekon2016}.


Throughout the literature, the 3D cosmic web structure is identified observationally by applying a variety of structure-finding algorithms \citep{Libeskind_2017} to large-scale spectroscopic galaxy surveys such as the SDSS, from which the 3D positions of galaxies can be extracted. By analysing the distribution of the galaxies in these surveys, it is also possible to estimate the density of matter in different regions of space~\citep{tempel2014detecting}, or to measure proxy statistics such as distance to nearest neighbour. However, methods based on galaxy catalogues suffer from survey incompleteness and uncertainty in galaxy distances. The most desirable way to define the cosmic web would be to use the underlying dark matter density field. Modern dark matter-only cosmological simulations can be used to calculate the cosmic web very precisely~\citep{angulo2022large}. However, most $N$-body simulations use initial conditions with random phases of density fluctuations in the early Universe, producing structure that resembles the actual Universe only statistically.

In this work we instead use \textit{constrained} simulations, which enable the encoding of both the amplitude and phases of the primordial density perturbations, giving the full 3D density and large-scale structure fields. Other key projects using constrained simulations include the ELUCID project~\citep{wang2016elucid,2018ApJ...860...30Y,2022MNRAS.517.3579Z,xu2023connectionsdssgalaxieselucid}, which investigates the relationship between constrained simulations and the SDSS, and the CLUES collaboration~\citep{2010arXiv1005.2687G,2016MNRAS.455.2078S} which aims to reconstruct the local universe and its clusters. The CLUES project uses the Weiner-filter method to reconstruct the mean field, but does not provide uncertainty quantification. Here, we adopt the \emph{Constrained Simulations in} \texttt{BORG} (\texttt{CSiBORG}) suite ~\citep{2021PhRvD.103b3523B,10.1093/mnras/stac2407,Desmond_2022,Bartlett_2022,kostić2023evidencepdwavedark,Stiskalek_2024, 2025arXiv250200121S}, based on the Bayesian Origin Reconstruction from Galaxies (\texttt{BORG}) algorithm~\citep{2013MNRAS.432..894J,2015JCAP...01..036J,2015PhDT.......367L,2016MNRAS.455.3169L,Jasche_2019,2024MNRAS.527.1244S,2024MNRAS.535.1258D}. The \texttt{BORG} algorithm is a Bayesian forward modelling algorithm that works as follows: it evolves the dark matter density field from initial conditions with some set of phases. The dark matter field is then populated with galaxies according to some galaxy bias model (relating dark matter and galaxy distribution), and a selection function corresponding to the relevant survey applied. The model galaxy field is then compared to observations through a likelihood, and the initial phases constrained through Bayesian inference. These phases are then used as initial conditions in \texttt{CSiBORG}, leading to a suite of 101 $N$-body simulations, corresponding to samples from the posterior. Hence, the \texttt{BORG} algorithm ensures that the large-scale structure in the \texttt{CSiBORG} simulations is positioned such that it matches the actual distribution of cosmic web structure in the Universe~\citep{Jasche_2019}.

In this study, we define a galaxy's cosmic web environment in two ways: the first using the dark matter density at the position of the galaxy as given by \texttt{CSiBORG} (smoothed on some scale), and the second by calculating the distance of the galaxy from the filaments and nodes of the cosmic web, which are found by applying the Discrete Persistence Structure Extractor algorithm (\texttt{DisPerSE};~\citealt{sousbie2011persistent, sousbie2011persistent2}) directly to the \texttt{CSiBORG} dark matter haloes. It should be noted that, in our first definition, galaxies are not assumed to lie at halo centres; the dark matter density is evaluated at the observed galaxy position in the smoothed \texttt{CSiBORG} field. By using constrained simulations to quantify cosmic web environment, we aim to measure the strength of correlations between a galaxy's environment and the properties of its gas and stars, which will constrain the relative importance of different galaxy formation processes. We aim to differentiate between those effects caused by the position of a galaxy relative to the 3D cosmic web structure, and those effects caused by the local dark matter density. Importantly, the fact that \texttt{CSiBORG} is a suite of simulations spanning the posterior of the \texttt{BORG} inference, means that we are able to propagate uncertainties in our knowledge of the dark matter field of the local Universe into our galaxy--environment correlation statistics. This affords, for the first time, a quantitative assessment of their statistical significance.

This paper is structured as follows: Section~\ref{sec:data} details the data used, including \texttt{CSiBORG}, NSA, and ALFALFA. Section~\ref{sec:method} details the methodology used to find correlations, define the structure of the cosmic web, smooth the \texttt{CSiBORG} fields and define quenching. Section~\ref{sec:results} presents the results and Section~\ref{sec:discussion} discusses their interpretation and comparison with the literature. Finally, Section~\ref{sec:conclusion} concludes with a summary of our findings. 

\section{Observed and Simulated Data}\label{sec:data}

\begin{table*}
  \centering
  \begin{tabularx}{\linewidth}{llXr} 
    \toprule 
    Quantity & Units & Description & Source Catalogue \\
    \toprule 
    $M_\star$ & $M_{\odot}$ & Total stellar mass of a galaxy. & NSA \\
    \midrule
    Colour & Magnitudes & $u-r$ band. & NSA \\
    \midrule
    S\'ersic index & - & 2D S\'ersic index, quantifying the concentration of a galaxy's light profile. & NSA\\
    \midrule
    Metallicity & - & Metallicity from k-correction fit. & NSA\\
    \midrule
    Size & Arcsec & Angular effective radius (Petrosian 50 per cent light radius).  & NSA\\
    \midrule
   $M_{\rm HI} / {M_\star}$ & - & The neutral hydrogen mass fraction of a galaxy, i.e., the total mass of \HI{} in a galaxy divided by its total stellar mass.  & ALFALFA \\
   \midrule
    SFR & $M_{\odot}/\mathrm{year}$ & The instantaneous star formation rate of a galaxy. & MPA/JHU \\
    \midrule
    sSFR & $1/\mathrm{year}$ & The instantaneous specific star formation rate of a galaxy (SFR/$M_\star$). & MPA/JHU \\
    \midrule
    $\rho_{\sigma}$
    where $\sigma = 0,2,4,8,16$ & $h^2 M_{\odot}/\rm{kpc}^3$ & The dark matter density field smoothed on a scale $\sigma$ in units of $\mathrm{Mpc} /h$. & \texttt{CSiBORG}\\
    \bottomrule
    \end{tabularx}
    \caption{Summary of quantities used in this work, including galaxy properties which we correlate with their environment, and the dark matter density from~\texttt{CSiBORG}. $\rho_0$ denotes the unsmoothed density field, which has a spatial resolution of 0.66 Mpc/$h$.}
    \label{tab:galaxy_properties}
\end{table*}

We use simulation data from \texttt{CSiBORG}, and galaxy properties from the NSA and ALFALFA surveys. The galaxy star formation rates are obtained from the Max Planck Institute for Astrophysics/John Hopkins University (MPA/JHU) survey. The properties of galaxies used throughout this report are summarised in Table~\ref{tab:galaxy_properties}. The effective volume and number of galaxies of each sample of data can be found in Table~\ref{tab:sample_properties}.

\begin{figure*}
    \centering
    \includegraphics[width=\textwidth]{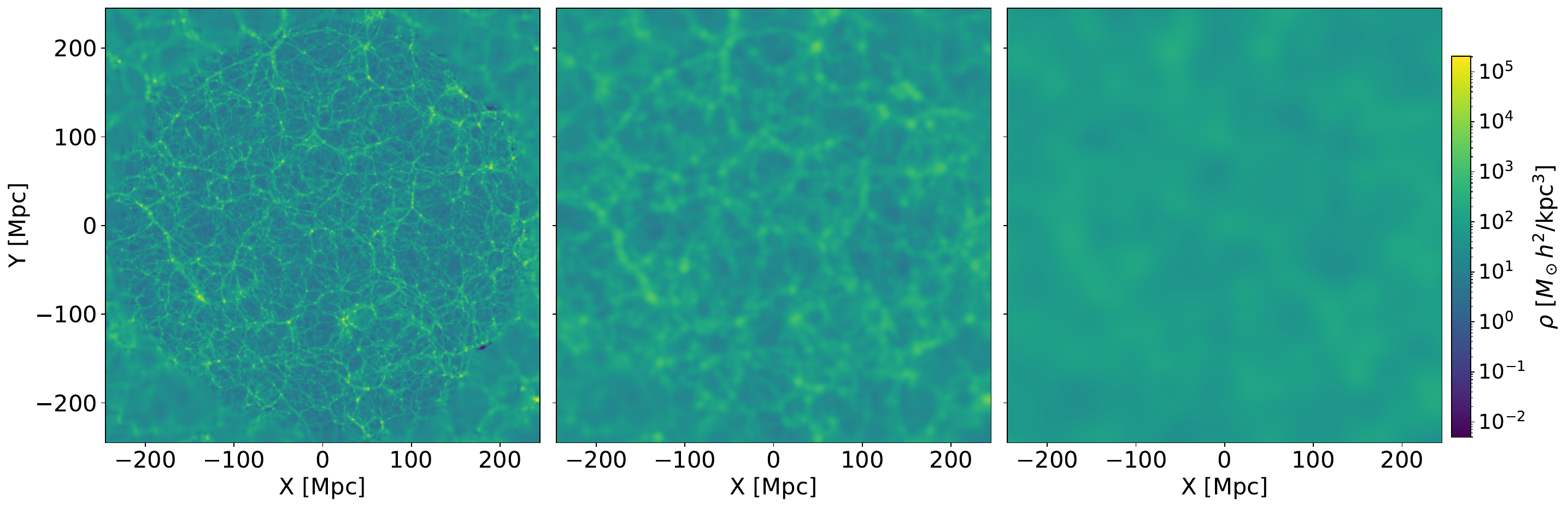}
    \caption{The dark matter distribution from a single realisation of \texttt{CSiBORG}, shown without and then with Gaussian smoothing of standard deviation $\sigma = 2$ and $8$ Mpc/$h$ (middle and right panel). Each panel represents a 2D slice through the Z-axis of the simulation box, centred on the Milky Way. The high-completeness central region of radius 155 Mpc/$h$ is visible. As smoothing increases the smaller scale structures are smoothed out. See Section~\ref{sec:csiborg} for further details of \texttt{CSiBORG}, and Section \ref{sec:density_stats} for further details of the smoothing methodology.}
    \label{density_diff_smoothing}
\end{figure*}

\subsection{Simulation Data}\label{sec:csiborg}

Introduced in~\citet{2021PhRvD.103b3523B}, the \texttt{CSiBORG} suite comprises 101 $N$-body simulations. Spanning a three-dimensional box with dimensions of $677.7$ Mpc/$h$, centred on the Milky Way, these simulations use initial conditions inferred from the Bayesian Origin Reconstruction from Galaxies algorithm (\texttt{BORG;}~\citealt{2013MNRAS.432..894J, 2015JCAP...01..036J,2015PhDT.......367L,Jasche_2019,2024MNRAS.535.1258D,2024MNRAS.527.1244S}) applied to the 2M\texttt{++} galaxy survey~\citep{Lavaux_2011}, which has a high completeness within a spherical region of radius $155$~Mpc/$h$ centred on the Milky Way. Specifically, this version of~\texttt{CSiBORG} is based on the initial conditions presented in~\cite{Jasche_2019}.

The 2M\texttt{++} reconstruction encompasses a cubic volume with a side length of $677.7$ Mpc/$h$ divided into $256^3$ voxels, resulting in a spatial resolution of $2.65$ Mpc/$h$. Similarly, \texttt{CSiBORG} features a high-resolution sphere centred on the Milky Way, with a radius of $155$ Mpc/$h$. In this region, initial conditions are augmented with white noise to account for random fluctuations on sub-constraint scales. The spatial distribution is defined on a grid measuring $2048^3$, yielding an initial inter-particle spacing of $0.33$ Mpc/$h$. Surrounding this high-resolution region, there is a spherical buffer zone with a width of $10$ Mpc/$h$, ensuring a smooth transition to the base \texttt{BORG} resolution. Figure~\ref{density_diff_smoothing} shows a slice of the \texttt{CSiBORG} box at different degrees of Gaussian smoothing (see Section~\ref{sec:density_stats}). The three plots are zoomed in on the central region of high completeness. The left panel clearly shows a cross section of the central spherical high-completeness region.

Dark matter haloes are identified as in~\citet{Stiskalek_2024}: within the central high-completeness region, the friends-of-friends halo finder (\texttt{FOF};~\citealt{1985ApJ...292..371D}) is used, with a linking-length parameter of $b=0.2$. The \texttt{FOF} algorithm connects particles within a distance $b$ times the mean particle separation. We require a minimum of 100 particles for a halo, corresponding to a minimum halo mass of $3.09 \times 10^{11} M_\odot h^{-1}$.

The dark matter density field is reconstructed from $z = 0$ particle snapshots. To convert them into a continuous density field, we employ the smoothed-particle hydrodynamics method \citep{1992ARA&A..30..543M,2007MNRAS.375..348C}. As outlined in Section IV.B.1 of \citet{Bartlett_2022}, a minimum of 32 neighbouring particles is required to smooth over. The resulting density field is then mapped onto a grid with a resolution of $0.7~\mathrm{Mpc} / h$. Both \texttt{BORG} and \texttt{CSiBORG} adopt the following cosmological parameters: $T_{\rm{CMB}} = 2.738$ K, $\Omega_m = 0.307$, $\Omega_{\Lambda} = 0.693$, $\Omega_b = 0.04825$, $H_0 = 70.05$ km/s/Mpc, $\sigma_8 = 0.8288$ and $n = 0.9611$. Lastly, for the purpose of this work we also use $20$ \texttt{CSiBORG}-like simulations but with random initial conditions to produce density fields that are uncorrelated with the observed galaxy properties.

\subsection{Nasa-Sloan Atlas}\label{sec:NSA_data}

We use the NSA survey\footnote{\url{http://nsatlas.org/data}} to obtain galaxy properties such as the stellar mass ($M_\star$), colour ($u-r$), S\'ersic index, size, and metallicity. A summary of these properties is provided in Table~\ref{tab:galaxy_properties}. However, it should be noted that the correlations with metallicity and stellar mass, presented in Section~\ref{sec:results}, should be used with caution, as stated explicitly in the NSA database.

The NSA database contains images and parameters of local galaxies, based on SDSS DR8~\citep{2011ApJS..193...29A} and the Galaxy Evolution Explorer survey (GALEX;~\citealt{martin2005galaxy}). We used the \texttt{nsa\_v1\_0\_1} data\footnote{\url{https://www.sdss4.org/dr17/manga/manga-target-selection/nsa/}}, which has a redshift range out to $z=0.15$ and contains approximately \num{640000} galaxies.

The distribution of these galaxies is illustrated in the right hand panel in Figure~\ref{mollweide_plots}. After restricting the NSA sample to the region covered by 2M\texttt{++}, which was used to constrain the \texttt{CSiBORG} density field, we obtain a total of approximately \num{90000} galaxies at a redshift of $z \leq 0.05$.

We obtain data for the star formation rate (SFR) and specific star formation rate (sSFR) of galaxies from MPA-JHU catalogue\footnote{\url{https://www.sdss4.org/dr17/spectro/galaxy_mpajhu/}}, which is based on \citet{2004ApJ...613..898T, 2003MNRAS.341...33K,2004MNRAS.351.1151B}. This catalogue is part of data release 8 of the SDSS, and provides derived galaxy properties based on spectral measurements.

\subsection{ALFALFA}\label{sec:alfalfa}

\begin{figure*}
    \centering
    \includegraphics[width=\textwidth]{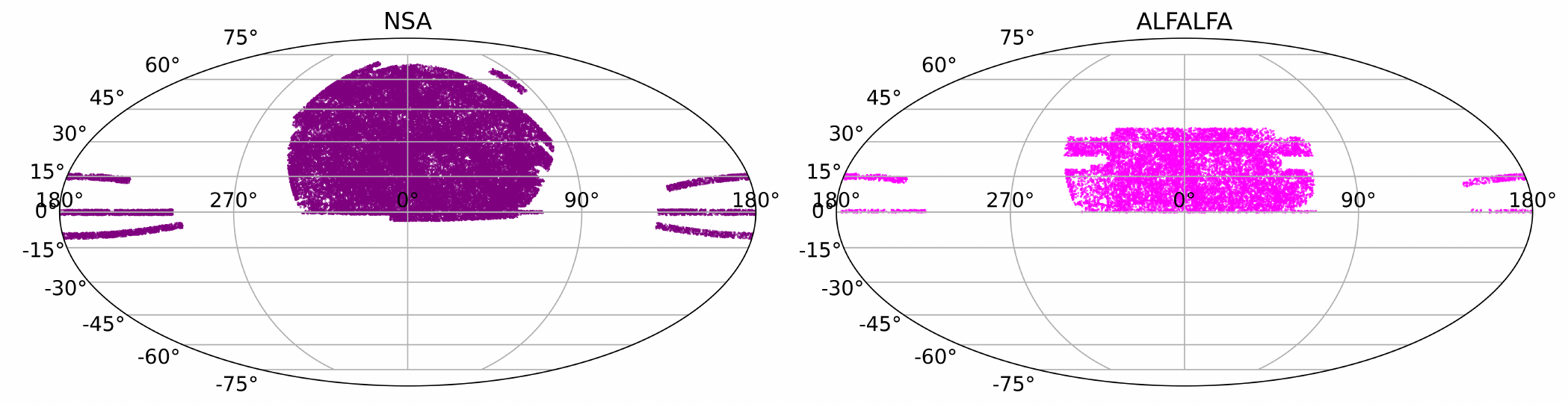}
    \caption{The distribution of NSA galaxies (left) and ALFALFA galaxies (right), shown on Mollweide projections of right ascension and declination. The NSA catalogue, based on optical data, contains information for 640,000 galaxies out to a redshift of 0.15. The ALFALFA catalogue provides \HI{} 21 cm line measurements for 31,500 galaxies out to a redshift of 0.06, covering nearly 7000 deg$^2$ of high galactic latitude sky. See Section~\ref{sec:NSA_data} and \ref{sec:alfalfa} for further information on the NSA and ALFALFA surveys respectively.}
    \label{mollweide_plots}
\end{figure*}

The ALFALFA\footnote{\url{http://egg.astro.cornell.edu/alfalfa/data/index.php}} survey maps nearly $7000~\rm{deg}^2$ of high galactic latitude sky using the Arecibo telescope, obtaining \HI{} 21 cm line measurements for \num{31500} galaxies out to a redshift of $0.06$ ~\citep{haynes2018arecibo}. We use the ALFALFA galaxy \HI{} masses.

We use an NSA and ALFALFA cross-match to obtain the $M_\star$, colour, SFR, S\'ersic index, size and \HI{} corresponding to ALFALFA galaxies. We match the NSA and ALFALFA catalogues by using a method following~\citet{2021MNRAS.506.3205S}, exploiting the partial overlap between ALFALFA and SDSS. This method matches the optical counterpart position to the sources in the NSA using an on-sky angle tolerance and line-of-sight tolerance of 5~arcsec and 10~Mpc respectively. This constitutes a strict criteria ensuring a low likelihood of mismatches and high sample purity. This cross-match retains approximately \num{22000} galaxies.  Due to the flux-density limit of ALFALFA, there is a relatively steep Malmquist bias with redshift for the HI{}-detected galaxies. Furthermore, the galaxies in this ALFALFA-NSA dataset are  bluer in colour than those from the general NSA galaxy sample, with higher star formation rates. This is expected, given the selection of galaxies rich in \HI{}. Consistent with this, we also find in this sample a clear correlation between star formation rate and $M_{\HI}$. The sky distribution of galaxies can be seen in the left hand plot of Figure~\ref{mollweide_plots}.

\begin{table}
  \centering
  \begin{tabularx}{\linewidth}{XXX} 
    \toprule 
    Sample & $N_{\rm gal}$&$V_{\rm eff}$$[10^6$ $\mathrm{Mpc}^3]$\\ 
    \toprule 
    NSA &  641,409 & -- \\
    \midrule
    NSA-CSiBORG & 89,328 & 2.2 \\
    \midrule
    NSA-CSiBORG-MPA/JHU & 81,281 & 2.0 \\
    \midrule
    ALFALFA & 31,500 & -- \\
    \midrule
    NSA-CSiBORG-ALFALFA & 22,478 & 0.56 \\
    \midrule
    NSA-CSiBORG-ALFALFA-MPA/JHU & 16,276 & 0.40  \\
    \bottomrule
    \end{tabularx}
    \caption{The number of galaxies ($N_{\rm gal}$) and effective volume ($V_{\rm eff}$) for each sample in $10^6$ Mpc$^3$, after cross-matching datasets and restricting to the CSiBORG high-completeness region. The effective volume was calculated using a voxel-based method applied to the 3D galaxy positions of the NSA-CSiBORG sample, and then inferred for the subsequent samples by assuming ratio of galaxies lost approximately corresponds to the fraction of volume lost. }
    \label{tab:sample_properties}
\end{table}

\section{Methodology}\label{sec:method}

\subsection{Density Statistics}
\label{sec:density_stats}

We begin by defining statistics of the dark matter density field, which we then correlate with galaxy properties. We do this as follows.  We conduct our analysis using 101 realisations of the \texttt{CSiBORG} simulations from which we obtain the dark matter density fields \p{}. For every realisation, we use the ``native'' SPH density field ($\rho_0$) along with four realisations of the field smoothed at different scales ($\rho_d$ for $d = 2,\,4,\,8,\,16~\mathrm{Mpc} / h$). We choose these degrees of smoothing in order to provide insights across a large range of spatial scales. We then correlate these with
the galaxy properties in Table~\ref{smoothing_equation}.
This allows us to investigate the effect of different spatial scales on various astrophysical processes. The Gaussian smoothing operates such that for a function $\rho(\mathbf{r})$, then
\begin{equation}
\rho_{\text{smoothed}}(\mathbf{r}) = \dfrac{1}{\sigma^3(2\pi)^{\frac{3}{2}}}\int_{\mathcal{R}} \rho_{\text{\rm DM}}(\mathbf{r'}) \exp{\left({\dfrac{|\mathbf{r}-\mathbf{r'}|^2}{2\sigma^2}}\right)} d\mathbf{r'},
\label{smoothing_equation}
\end{equation}
where $\sigma$ is the standard deviation of the Gaussian smoothing kernel in units of Mpc/$h$. The smoothed dark matter density field is rendered on a $1024^3$ grid such that the baseline of ``no additional smoothing'' has a spatial resolution of $0.66$ Mpc/$h$. This approximately corresponds to the typical scale of a cluster-sized halo~\citep{bahcall1999clusters}. Note however that the constraints on the initial conditions are on a 2.6 Mpc/$h$ scale, and hence below this there is limited information on the actual density field. Figure~\ref{density_diff_smoothing} shows a 2D slice of a single \texttt{CSiBORG} realisation of the dark matter density field for $\rho_0$, $\rho_2$, and $\rho_8$. The effect of Gaussian smoothing can be seen on these plots, which show how finer features become more blurred as smoothing increases.

Estimating the dark matter density at the galaxy position should take into account its distance uncertainty. The NSA catalogue provides distances calculated using the Willick flow model~\citep{Willick_1997}, however no uncertainty is provided. The expected typical distance uncertainty due to small-scale velocities not accounted for by the model, is $\sim300$~km/s~\citep{2025arXiv250200121S}, which corresponds to a distance uncertainty of $\sim3$~Mpc/$h$. Therefore even our smallest smoothing scale should smooth out variations in the dark matter density comparably to the distance uncertainty (and the 2.6~Mpc/$h$ scale of the initial condition constraints).

\subsection{Cosmic web statistics}\label{sec:cosmic_web_stats}

\begin{figure*}
    \includegraphics[width=\linewidth]{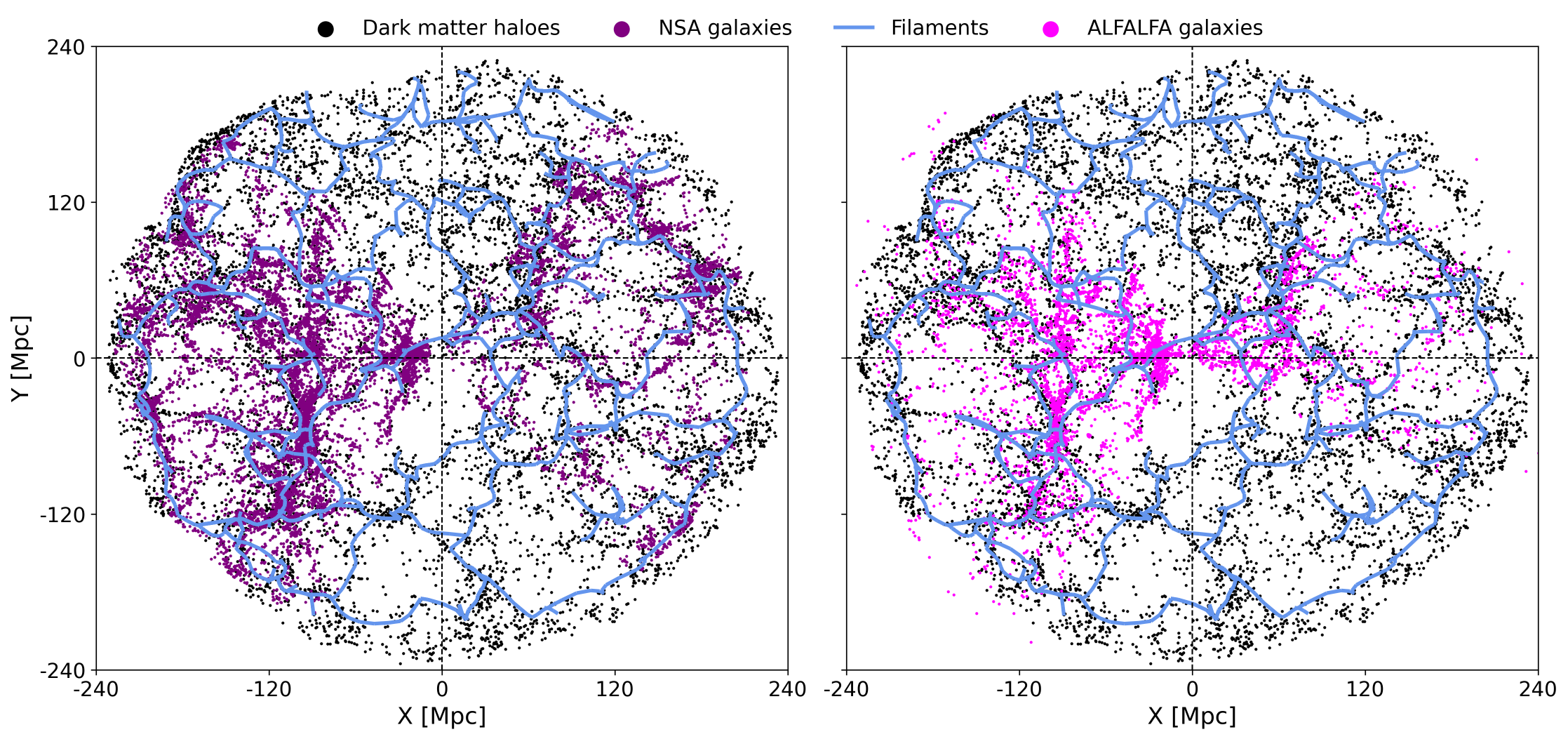}
    \caption{Distribution of dark matter haloes (black) derived from \texttt{CSiBORG}, overlaid with NSA galaxies (purple; left panel), ALFALFA galaxies (magenta; right panel) and cosmic web filaments (blue), shown in a 15 Mpc slice along the z-axis centred on the Milky Way. The cosmic web filaments are calculated using \texttt{DisPerSE}, applied to the \texttt{CSiBORG} haloes. NSA and ALFALFA galaxies predominantly cluster along the filamentary structures, with some galaxies distributed within the voids of the cosmic web. See Section~\ref{sec:csiborg} for details on the \texttt{CSiBORG} haloes, and Section~\ref{sec:cosmic_web_stats} for further details on the \texttt{DisPerSE} methodology and filament extraction.
    }
    \label{fig:cosmic_web}
\end{figure*}

To define the cosmic web we use the Discrete Persistence Structure Extractor\footnote{\url{https://www2.iap.fr/users/sousbie/web/html/indexd41d.html}} algorithm (\texttt{DisPerSE};~\citealt{sousbie2011persistent,sousbie2011persistent2}), which uses discrete Morse theory~\citep{milner1963discrete} to detect topological features: maxima, minima and saddle points. These features can then be used to convert the density field into the structural elements of the cosmic web: nodes, filaments, sheets, and voids. 

\texttt{DisPerSE} can be applied to any discrete point distribution and is most commonly used in observational studies on galaxy catalogues. However, our use of constrained simulations enables us to apply it directly to the dark matter at either the particle or halo level, sidestepping issues associated with galaxy bias. We choose to apply it to the halo catalogues rather than particle data because 
this approach has been tested and calibrated by \citet{galárragaespinosa2024evolutioncosmicfilamentsmtng}. In that work, the \texttt{DisPerSE} parameters were calibrated for halo catalogues to maximise the purity and completeness of associating nodes with the positions of massive haloes in non-constrained simulations. We outline their method and parameter choices below. We apply \texttt{DisPerSE} to all 101 realisations of \texttt{CSiBORG} dark matter haloes, along with 20 unconstrained simulations to allow us to determine the statistical significance of our results (see Section~\ref{sec:results}).

The \texttt{DisPerSE} algorithm starts by constructing a 3-dimensional Delaunay tessellation field, using the Delaunay Tessellation Field Estimator (DTFE; 
\citealt{schaap2000continuousfieldsdiscretesamples}). The DTFE density field is estimated directly from halo positions without applying halo-mass weights, as in~\citet{galárragaespinosa2024evolutioncosmicfilamentsmtng}. We use periodic boundary conditions to estimate the particle density field. We then apply the ``netconv'' smoothing function to this density field, taking the smoothing threshold value as $1\sigma$. We then apply the ``mse'' function, which is the ``manifold skeleton extractor'' and the primary function of \texttt{DisPerSE}. This identifies the critical points of the density field and requires the user to input a persistence threshold, which we take to be 2$\sigma$. This threshold is important as if it is too low, it may lead \texttt{DisPerSE} to map noise as filaments, or if too high, \texttt{DisPerSE} may miss the finer structures.

The values that we take for smoothing and persistence threshold are consistent with~\citet{galárragaespinosa2024evolutioncosmicfilamentsmtng}, who look for the the optimal smoothing and persistence threshold parameters that recover the most accurate cosmic web features while minimising noise. We then apply the ``skelconv'' function to convert the outputs into a readable format, with critical points indicating when the gradients of the manifold are equal to zero. The critical points are maxima, minima, and saddle points. The filament arms are given by connecting a maxima to a saddle point and are made up of many small segments. The filament distribution overlaid on one realisation of the dark matter haloes, alongside the observational data, is shown in Figure~\ref{fig:cosmic_web}.

With the filament locations identified, we calculate the midpoint of each segment, and compare these with the galaxy coordinates from the NSA database, in order to compile a list of the ``distance to the nearest filament'' for every galaxy, which we hereafter refer to as \df{}. As the average segment size is about 1~Mpc, the error from the discretisation of the filament is negligible. The process produces 101 separate sets of \df{}, each corresponding to a realisation of the \texttt{CSiBORG} dark matter haloes. Additionally we obtain 20 sets of \df{} corresponding to the unconstrained simulations.

\subsection{Correlation Statistics}\label{sec:corr_stats}

We use two methods to calculate the correlation between data sets: a direct Spearman correlation between two properties (``direct correlation''), and a partial Spearman correlation conditioned on $M_\star$, \p{}, or \df{} (``partial correlation'').

The Spearman rank correlation coefficient quantifies whether a relation between two variables can be described by a monotonic function. Unlike the Pearson correlation coefficient, which measures linear relationships, the Spearman correlation coefficient takes into account the ranks of the data points rather than their raw values. Therefore, it is less sensitive to outliers, since outliers have a smaller effect on the ranking of the data. The Spearman correlation coefficient is calculated by considering $n$ pairs of observations from two distributions, with the observations in each distribution ranked from smallest to largest~\citep{spearman_eqn}. Then, if $d_i$ is the difference between the ranks of each observation, the Spearman rank correlation coefficient, $r_S$, is given by
\begin{equation}\label{spearman_equation}
    r_S = 1 - \dfrac{6\sum d_i^2}{n(n^2-1)}.
\end{equation}

We also use a partial correlation method to quantify the relationship between a galaxy property and either \p{} or \df{}, while controlling for a third property: $M_\star$, \p{} or \df{}. This is achieved by fitting a locally weighted scatter plot function (LOWESS\footnote{\url{https://www.statsmodels.org/dev/generated/statsmodels.nonparametric.smoothers_lowess.lowess.html}}; \citealt{Cleveland01121979}) between each variable of interest ($A$ and $B$) and the third property ($C$, the condition). LOWESS fits a generalised non-linear function to the data, by locally fitting a weighted least squares regression line. We use LOWESS as it handles outliers more reliably than a basic curve fit function, and allows for flexibility in the fit when handling complex data with a non-trivial relationship. We then compute the residuals from the fit between A and C, and then between B and C. The Spearman rank correlation coefficient is calculated between the two sets of residuals, providing the partial correlation between A and B, conditioned on property $C$.

\begin{figure*}
    \includegraphics[width=0.9\textwidth]{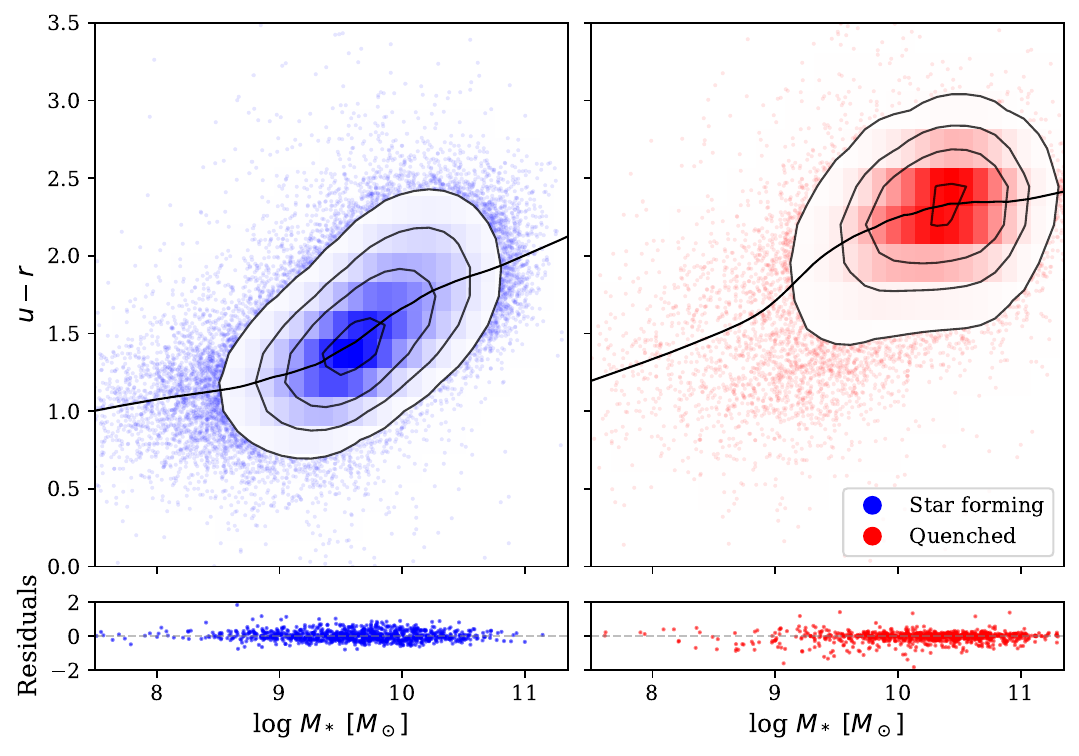}
    \caption{The relationship between colour ($u-r$) and stellar mass, with the solid black line showing the LOWESS fit. The bimodal distribution of colour has been separated into ``star-forming'' (left) and ``quenched'' (right). The contour lines show the points contained within $0.5\sigma$, $1\sigma$, $1.5\sigma$, and $2\sigma$, which represents 11.8$\%$, 39.3$\%$, 67.5$\%$, and 86.4$\%$ regions of each distribution, respectively.}
    \label{fig:colour_fit}
\end{figure*}

\begin{figure}
    \includegraphics[width=\linewidth]{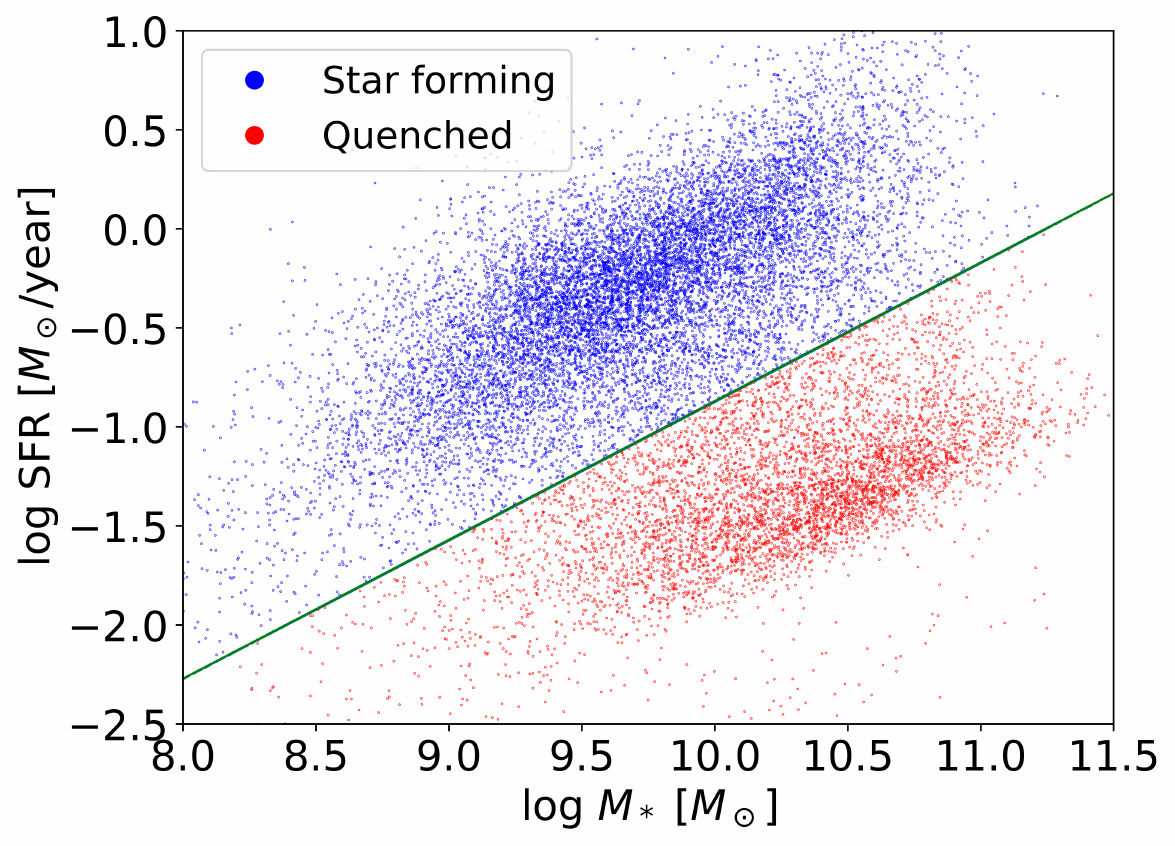}
    \caption{The bimodal distribution of the star formation rate of NSA galaxies (from the MPA/JHU catalogue), classifying galaxies as star-forming (blue; above the dividing line) and quenched (red; below the dividing line). We use Equation~\ref{eq:quenching} to divide the two populations. See Section \ref{sec:quenched_fraction} for further details on the methodology of defining a quenched galaxy.}
    \label{quenching_method}
\end{figure}

Where there is a bimodal relationship between a variable and stellar mass, we separate the two distributions. For example, with colour, SFR, and sSFR we separate ``quenched'' and ``star-forming'' galaxies in line with the process later described in Section~\ref{sec:quenched_fraction} and visible in Figure~\ref{quenching_method}.
We find the residuals of these distributions separately. Figure~\ref{fig:colour_fit} shows how the residuals were found for colour, after splitting the bimodal colour distribution as star-forming or quenched galaxies.

We apply the above procedure to extract the correlation coefficients between galaxy and environmental properties (\p{} and \df{}) for the 101 constrained simulations as well as the 20 unconstrained simulations. We find that for the unconstrained simulations, the distributions of correlation coefficients are approximately Gaussian and centred on zero. We find that the standard deviations of the distributions are much smaller for the constrained simulations than the unconstrained simulations, showing the uncertainty due to the BORG posterior is small. We perform bootstrapping with replacement to test the uncertainty due to the finite sample size of the galaxy surveys, but find it is negligible.

Given the above, we quantify the statistical significance of each correlation as the mean of the correlation coefficients from the constrained simulations divided by the standard deviation of the coefficients from the unconstrained simulations. This tests against the null hypothesis of no correlation between environment and any of the galaxy properties, which is necessarily the case in the unconstrained simulations.

\subsection{Quenched Fraction}\label{sec:quenched_fraction}

We also study the dependence of the quenched fraction of galaxies on \p{} and \df{}.
In Figure~\ref{quenching_method} we plot the star formation rate (from the MPA/JHU catalogue) against $M_\star$, which produces a bimodal distribution of galaxies. This can be split into galaxies on the ``main sequence,'' which we define as star-forming, and those below this main sequence band, which we define as quenched.

We define galaxies as star-forming or quenched with the following definition of the main sequence from~\citet{Whitaker_2012}:
 \begin{equation}
 \label{eq:quenching}
     \log_{10}(\text{SFR/M}_{\odot}~\text{yr}^{-1}) = \alpha(z)[\log_{10} (M_\star/\text{M}_{\odot})-10.5] + \beta(z),
 \end{equation}
where the slope $\alpha(z) = 0.70 - 0.13z$ and the normalisation factor $\beta(z) = 0.38 + 1.14z - 0.19 z^2$.
We then subtract 1 dex from $\log_{10}(\text{SFR})$ in order to find the line separating the star-forming and quenched sequences, as this is approximately at the minima between the star formation main sequence and the population of quenched galaxies.

\section{Results}\label{sec:results}

Here we present the results obtained for the relationship of galaxy properties (described in Table~\ref{tab:galaxy_properties}) with environmental density \p{} and distance from filament \df{}. For simplicity, we will henceforth refer to the ALFALFA$\times$NSA data as ``ALFALFA data''. For a direct comparison between the results presented here and the corresponding literature, please see Sections~\ref{sec:interpretation} and ~\ref{sec:litrev}.

\begin{figure}
    \includegraphics[width=\linewidth]{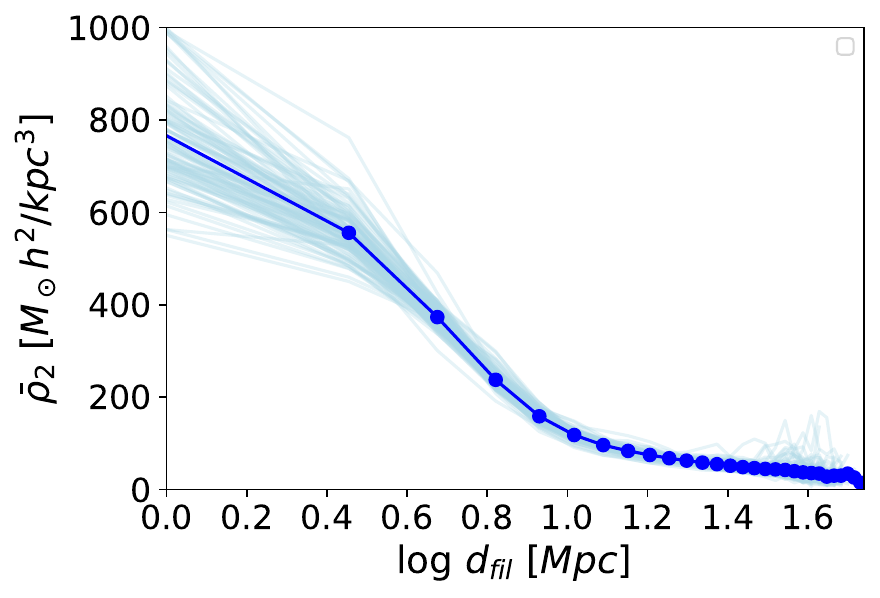}
    \caption{The relationship between the matter density $\rho_2$ and distance to nearest filament \df{} evaluated at the position of NSA galaxies. The thin lines are the relations inferred from individual \texttt{CSiBORG} realisations, whereas the bold blue is their mean. $\rho_2$ rapidly decreases with \df{}, affirming that the dark matter density is concentrated around the filament, and that there is a sharp density gradient surrounding the filament.
    }
    \label{fig:density_vs_distance}
\end{figure}

In Figure~\ref{fig:density_vs_distance} we show the relationship between dark matter density and distance from filament (inferred from \texttt{CSiBORG}) evaluated at the position of NSA galaxies. As expected, there is a strong anti-correlation between mean density and distance to filament. To fully investigate the relative effect of filament distance and density on galaxy properties, we study their correlations with \p{} and \df{} individually, as well as their correlation with \p{} conditioned on \df{} and vice-versa.

\subsection{NSA Correlations}

\begin{figure*}
    \includegraphics[width=\textwidth]{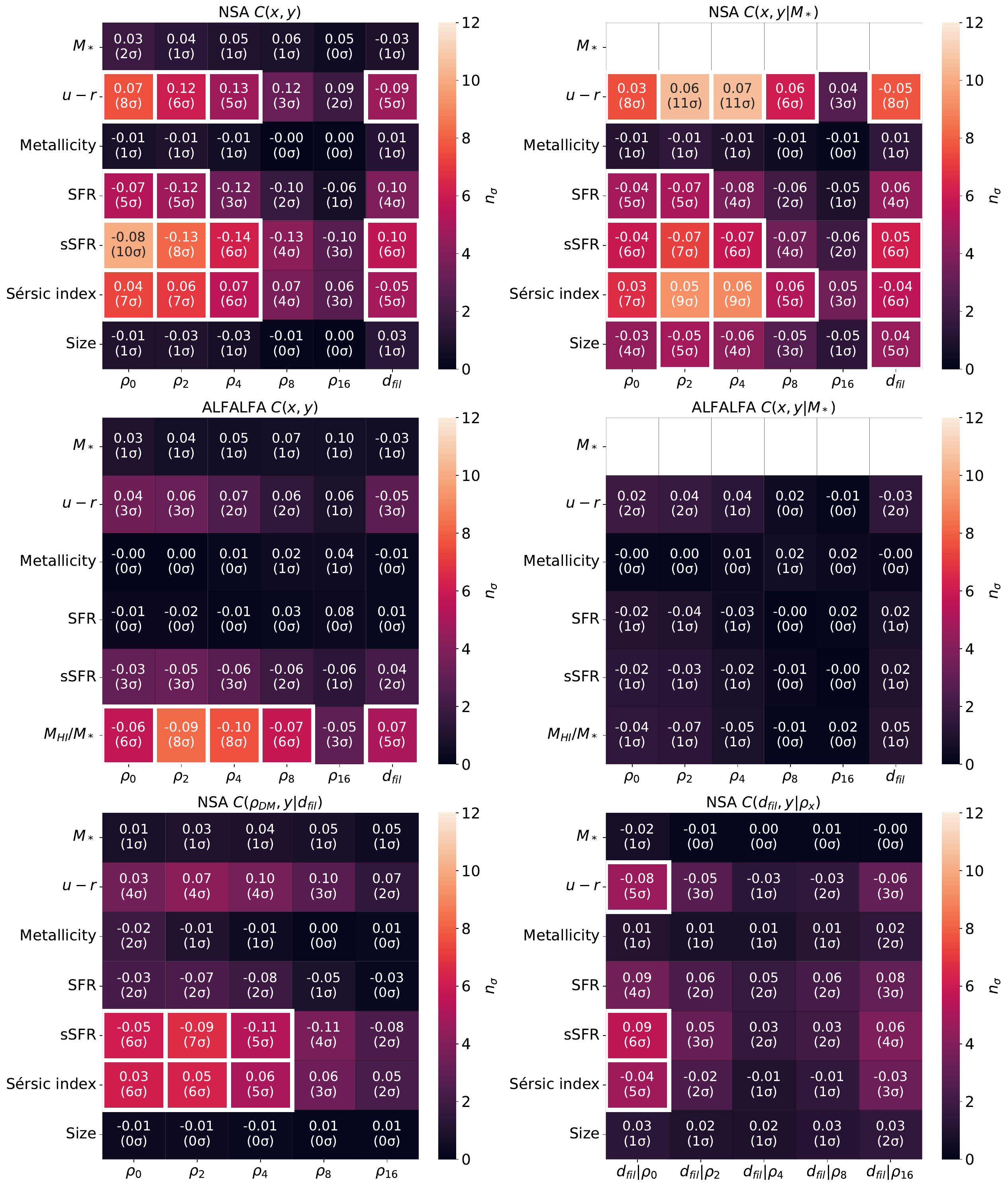}
    \caption{Correlation coefficients of galaxy properties with \p{} and \df{}, along with the associated uncertainty and statistical significance of the correlation given in brackets.} A white box indicates a correlation of at least $5\sigma$ statistical significance. The first row shows the direct correlation (left) and partial correlation conditioned on $M_\star$ (right) between NSA galaxy properties and \p{} or \df{}. The second row shows the same correlations for ALFALFA galaxies. The third row shows the partial correlation between NSA galaxy properties and \p{} conditioned on \df{} (left), and the partial correlation between NSA galaxy properties and \df{} conditioned on \p{} (right). Correlations of galaxy properties with the dark matter density are typically stronger than those with the filament distance. See Section~\ref{sec:results} and Section~\ref{sec:interpretation} for a description and interpretation of these results respectively.
    \label{fig:spearman_correlations_comparison}
\end{figure*}

\begin{figure*}
    \centering
    \begin{subfigure}[b]{0.49\linewidth}
        \includegraphics[width=\linewidth]{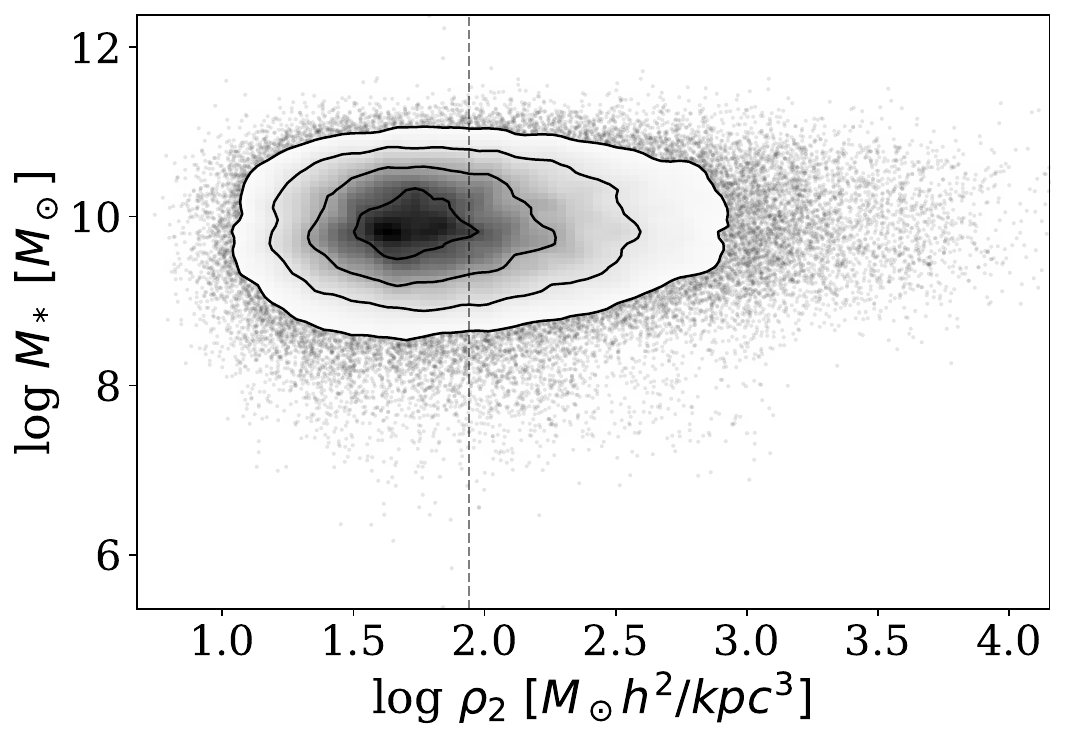}
        \label{fig:mass_vs_density}
    \end{subfigure}
    \hfill
    \begin{subfigure}[b]{0.49\linewidth}
        \includegraphics[width=\linewidth]{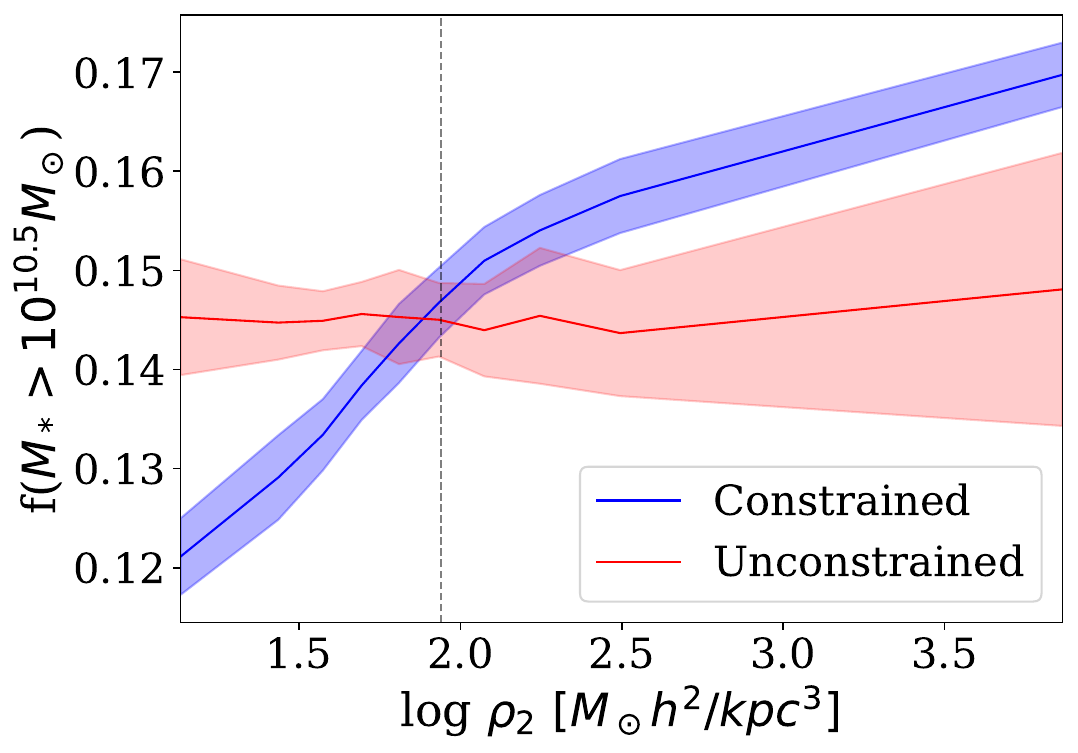}
        \label{fig:high_mass_fraction}
    \end{subfigure}
    \caption{The relationship between dark matter density $\rho_2$ and stellar mass $M_\star$ (left) and the fraction of high mass galaxies ($M_\star > 10^{10.5} M_\odot$) in bins of $\rho_2$ (right). The dashed vertical line on each plot shows the average matter density. The contour lines on the left panel are at $0.5\sigma$, $1\sigma$, $1.5\sigma$, and $2\sigma$, containing 11.8$\%$, 39.3$\%$, 67.5$\%$, and 86.4$\%$ of the points, respectively. The solid lines on the right panel show the mean fraction of high mass galaxies in density bins across 101 constrained (blue) simulations and 20  unconstrained (red) simulations, with the shaded regions representing the standard deviation across the respective set of simulations. The constrained simulations show a positive correlation between fraction of high mass galaxies and density, while the unconstrained simulations show no correlation and a very high standard deviation towards high densities.}
    \label{fig:density_and_mass_fraction}
\end{figure*}

The results for the Spearman correlations of galaxy properties taken from the NSA survey (see Table~\ref{tab:galaxy_properties}) with \p{} and \df{} are shown in the top and bottom rows of Figure~\ref{fig:spearman_correlations_comparison}. This figure shows the direct correlation between galaxy properties with \p{} and \df{} in the top left panel, the partial correlation conditioned on $M_\star$ between galaxy properties with \p{} and \df{} in the top right panel, the partial correlation between galaxy properties and \p{} conditioned on \df{} in the bottom left panel, and the partial correlation with \df{} conditioned on \p{} in the bottom right panel. We show the uncertainty, or `statistical significance' of each correlation in parentheses below the value of the correlation, with a description of how this was calculated in Section~\ref{sec:corr_stats}.


The direct correlations of both \p{} and \df{} with galaxy properties (colour, SFR, sSFR and S\'ersic index) are statistically significant. For \p{}, the correlation peaks around $\rho_0$ or $\rho_2$, generally showing stronger correlations than those for \df{}, with comparable or higher statistical significance (e.g. correlation coefficient of -0.13 with statistical significance $8\sigma$ for $\rho_2$ with sSFR, compared to 0.10 and statistical significance $6\sigma$ for \df{}). When conditioned on stellar mass, size displays stronger correlations with \p{} than the direct correlation. Notably, conditioning on \df{} yields stronger correlations than conditioning on \p{}.

There is a notably weak correlation between $M_\star$ and \p{} with a low statistical significance. We tested this by defining a set of high mass galaxies ($M_\star>10^{10.5}M_\odot$) and finding the fraction of high mass galaxies in bins of \p{}. The right panel of Figure~\ref{fig:density_and_mass_fraction} shows this fraction as a function of density: we can see that high mass galaxies tend to be found in denser environments. \citet{2025ApJ...988..280W} study the dependence of galaxy luminosity functions on environment, for different density bins. Their Figure 2 shows that at the faint end of the luminosity function, the number density of galaxies is nearly independent of environment. At the bright end, however, the number density is reduced in low-density environments compared to high-density regions. This supports the idea that high mass galaxies tend to reside in higher density environments, and suggests that low mass galaxies are scattered across all environmental densities. It is likely that the more scattered distribution of low mass galaxies washes out the correlation between $M_\star$ and \p{}, hence why we see a very low correlation between $M_\star$ and \p{} in Figure~\ref{fig:spearman_correlations_comparison}.

\subsection{ALFALFA correlations}\label{sec:alf_results}

Next we investigate the correlation between \p{} and \df{} with galaxy properties for the ALFALFA galaxies. In addition to the galaxy properties investigated for the NSA data, we also include the $M_{\HI}/M_\star$. The correlation results are presented in the middle row of Figure~\ref{fig:spearman_correlations_comparison}, with the direct correlation in the left panel, and the partial correlation conditioned on $M_\star$ in the right panel.

The direct correlation for the ALFALFA data reveals statistically significant negative correlations between $M_{\HI}/M_\star$ and \p{}, with the strongest correlation and peak statistical significance with $\rho_4$. The ALFALFA partial correlation conditioned on $M_\star$ reveals very weak correlations of low statistical significance across all galaxy properties. The highest statistical significance is $2\sigma$ for the correlations with colour. The ALFALFA partial correlations conditioned on $M_\star$ between any property and \df{} are weak and of low statistical significance.

The weak correlations displayed by the ALFALFA galaxy properties are likely due to the selection effects of the ALFALFA survey, which contains preferentially bluer, star-forming galaxies, as these tend to be \HI{} rich. This selection results in a narrower range of galaxy properties, reducing the observed correlation strengths. Additionally, the smaller sample size of the ALFALFA dataset likely contributes to the weak correlations and low statistical significance.

\subsection{Galaxy quenching}\label{sec:galaxy_quiescence}

\begin{figure*}
    \centering
    \includegraphics[width=\textwidth]{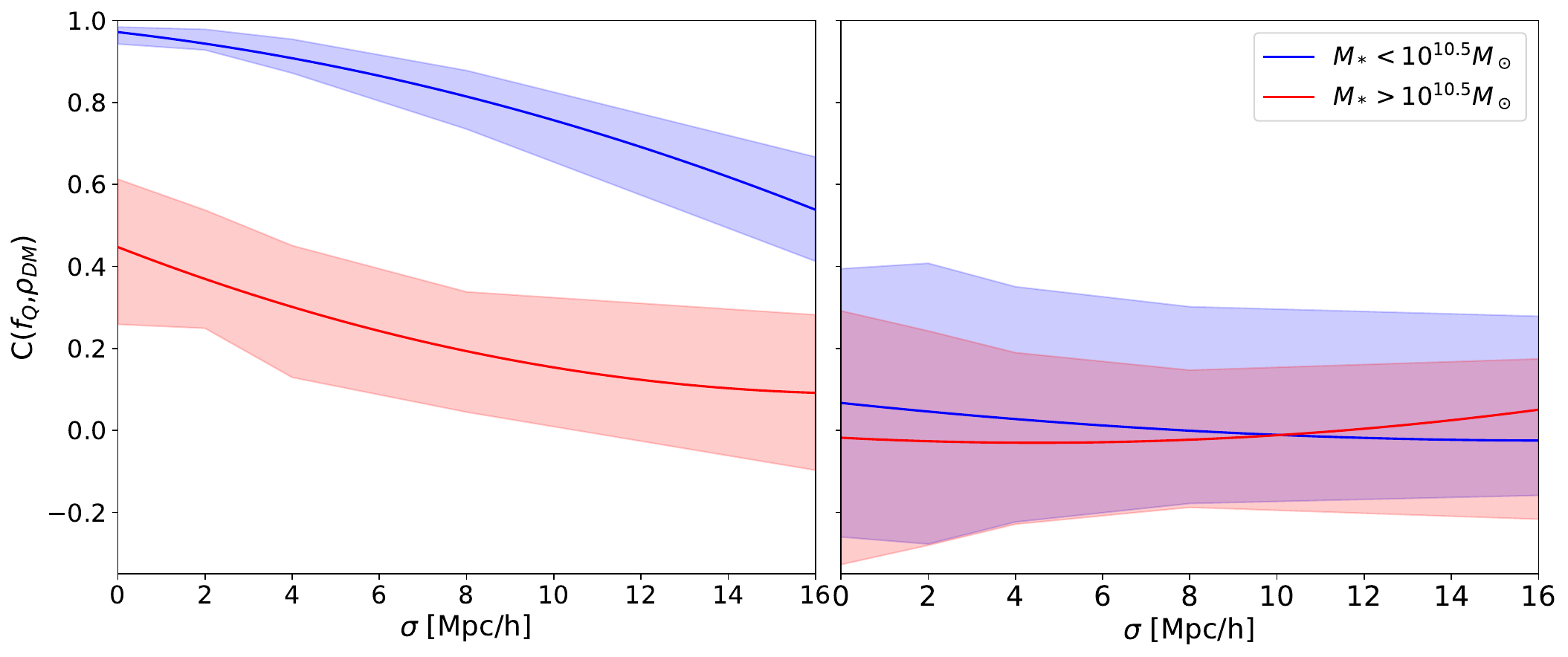}
    \caption{The mean correlation between the fraction of quenched galaxies $f_Q$ and the environmental density \p{} from constrained simulations (left) and unconstrained simulations (right) as a function of the Gaussian smoothing scale, $\sigma$. The solid lines and shading show the mean and $1\sigma$ uncertainty across the \texttt{CSiBORG} realisations, respectively. We separately show low mass (blue, $M_\star < 10^{10.5} M_\odot$) and high mass (red, $M_\star > 10^{10.5} M_\odot$) galaxies. Low mass galaxies exhibit a much stronger correlation with environment than high mass galaxies, suggesting that quenching in low mass galaxies is primarily driven by external mechanisms, while in high mass galaxies, internal mechanisms play a larger role. See Section~\ref{sec:galaxy_quiescence} and Section~\ref{sec:quenching_discussion} for further description and analysis of these results.}
    \label{quenching_trend}
\end{figure*}

We now investigate the relationship between the dark matter density, distance to nearest filament and the quenching of galaxies in the NSA catalogue. We test this separately for low and high mass galaxies, as previous studies suggest that quenching mechanisms can depend on galaxy mass~\citep{2014MNRAS.441..599B,quenching_paper,2025arXiv250100986Z}. Following~\citet{quenching_paper}, we define a high mass galaxy as $M_\star>10^{10.5} M_{\odot}$.

Figure~\ref{quenching_trend} shows the correlation between \p{} and the fraction of quenched galaxies across different smoothing scales. The solid lines represent the average correlation (mean of 101 realisations of \texttt{CSiBORG}) as a function of smoothing scale for high mass (red) and low mass (blue) galaxies. The shaded area shows the standard deviation across the 101 realisations.

The correlation between the fraction of quenched galaxies and \p{} is stronger for low mass galaxies than for high mass galaxies. For both high and low mass galaxies, the correlation decreases while smoothing scale increases. The standard deviation increases towards $\rho_{16}$ (particularly for the low mass galaxies), likely due to the fact that the result becomes increasingly randomised as the scale becomes less relevant. Following the split, we have $\sim$ 76,000 and $\sim$ 13,000 low- and high-mass galaxies, respectively, thus the standard deviation in the correlation coefficient of high mass galaxies is larger.

\begin{figure*}
    \centering
    \includegraphics[width=\textwidth]{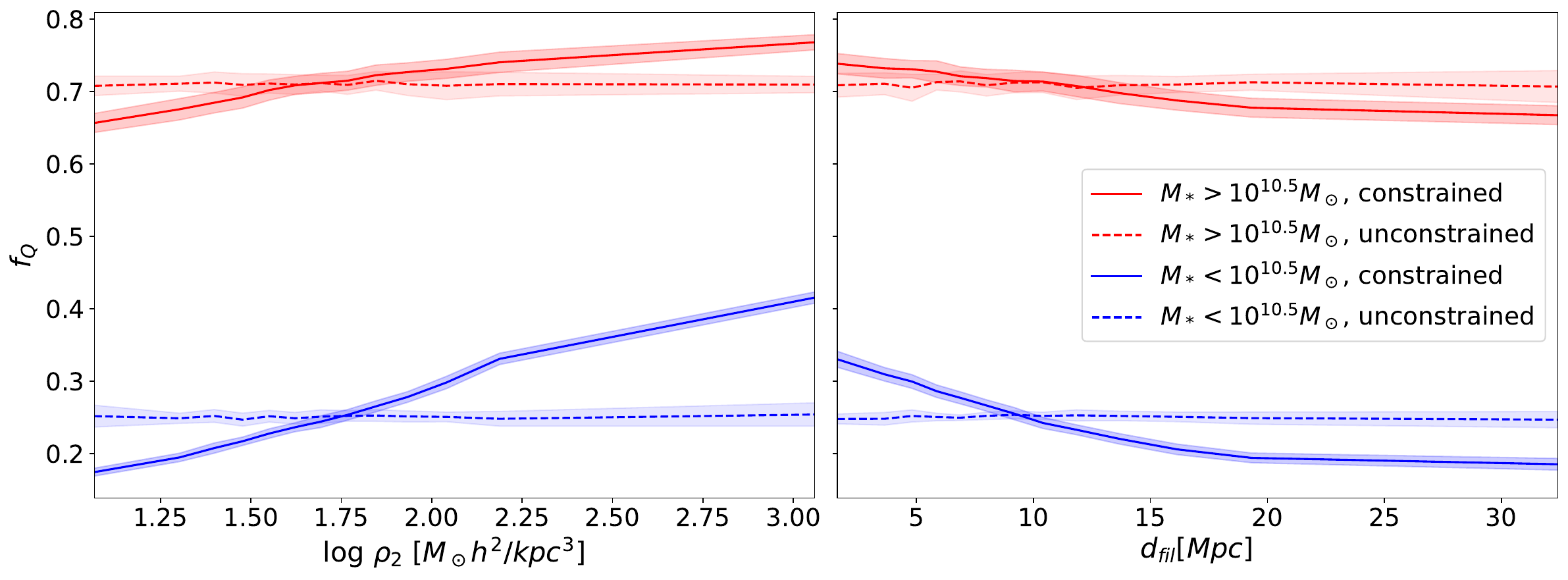}
    \caption{The fraction of quenched galaxies $f_Q$ across bins of $\rho_2$ (left) and bins of \df{} (right) for low mass (blue, $M_\star < 10^{10.5} M_\odot$) and high mass (red, $M_\star > 10^{10.5} M_\odot$) galaxies.
    The solid lines and shading show the mean correlation and its $1\sigma$ uncertainty, respectively, for constrained simulations. The dotted lines and shading show the equivalent for unconstrained simulations. $f_Q$ increases as $\rho_2$ increases, suggesting that greater environmental densities are associated with higher quenching rates. $f_Q$ decreases as \df{} increases, indicating that galaxies further from filaments are less likely to experience quenching. $f_Q$ begins to flatten at around $d_{\rm fil} = 15~\mathrm{Mpc}$, while $f_Q$ increases for each value of $\rho_2$. When compared to unconstrained simulations, the plots show that the relationship between $f_Q$ and environment is far stronger for low mass galaxies than for high mass galaxies. See Section~\ref{sec:quenching_discussion} for further interpretation of these results.}
    \label{fig:quenching_comparison}
\end{figure*}

Lastly, Figure~\ref{fig:quenching_comparison} shows how the fraction of quenched galaxies changes with \df{} and $\rho_2$. We choose $\rho_2$ for this analysis, as this scale produces the greatest statistical significance for the majority of the correlations in Figure~\ref{fig:spearman_correlations_comparison}. We see a stronger correlation between environment and quenched fraction for low mass galaxies than high mass galaxies, in line with Figure~\ref{quenching_trend}. We also see that while quenched fraction continues to increase with $\rho_2$, it begins to flatten at around 15 Mpc for \df{}.

\section{Discussion}\label{sec:discussion}

\subsection{Interpretation of results}\label{sec:interpretation}

An advantage of \texttt{CSiBORG} is that the multiple realisations correspond to the uncertainty with which we know the density field. By comparing the correlation strength of each realisation of the constrained simulations we can see this uncertainty is subdominant. This implies a low uncertainty due to the \texttt{BORG} posterior. Additionally, bootstrapping galaxies within the sample shows that the uncertainty due to sample size is minimal. The dominant source of uncertainty arises from comparing the correlation from constrained simulations to that from unconstrained simulations, a process described in Section~\ref{sec:corr_stats}.

\subsubsection{Star formation and \HI{} content}\label{sec:sfr_discussion}

We now consider the implications of the SFR and $M_{\HI}/M_\star$ correlations on models of galaxy formation. Both the NSA and ALFALFA correlation results imply that as \p{} increases, a galaxy is less likely to be star-forming, will have a lower $M_{\HI}/M_\star$, and will be redder in colour. In the literature, these trends are often attributed to effects such as ``gas heating'' (where the intergalactic medium around galaxies is heated by external sources like shocks or AGN feedback, preventing gas from cooling or forming stars, e.g.~\citealt{2007ARA&A..45..117M}), ``strangulation'' (the cessation of cold gas inflow onto a galaxy, e.g.~\citealt{2008ApJ...672L.103K}), or the effect of ``galaxy group interactions'' near filaments (for example ram pressure stripping, mergers, and tidal forces between galaxies which often lead to tidal stripping, e.g.~\citealt{1972ApJ...176....1G, rodiger2005ram, 2008MNRAS.383..593M}).

Consistent with the correlations for \p{}, the results for \df{} imply that galaxies closer to filaments have a lower $M_{\HI}/M_\star$ and are redder in colour. While of reasonable statistical significance, the direct correlation coefficients are weak. The partial correlations conditioned on $M_\star$ are of negligible statistical significance. Together, these results suggest that gas feeding galaxies from filaments is not strongly prevalent in the local Universe. Our results are qualitatively consistent with \citet{2018MNRAS.474..547K}, who find that at lower redshift, filaments are not expected to boost gas flow and therefore increase the quantity of gas or star formation rate in galaxies. However at high redshifts, the role of the cosmic web is expected to be more important, with gas flow from filaments into galaxies an important effect, as shown by \citet{ramsoy2021rivers}.

\subsubsection{Quenching}\label{sec:quenching_discussion}

The results presented on the relationship between fraction of quenched galaxies and \p{} align with~\citet{quenching_paper}, who argue that low mass galaxies experience quenching predominantly due to their environment. \citet{quenching_paper} link quenching in low mass galaxies to properties such as host halo mass and galaxy overdensity. On the other hand, they find that quenching in high mass galaxies is likely to be due to internal processes such as AGN feedback. Our results support this idea, showing that the quenching of high mass galaxies has a low dependence on environment. This is evident in Figure~\ref{quenching_trend}, where the correlation between \p{} and quenched fraction is far weaker for high mass galaxies than low mass galaxies. \citet{2018A&A...612A..31P} find that the relationship between AGN properties and environmental density inferred from \texttt{BORG} is more important than the cosmic web structure. However, an alternative possibility is that the most highly quenched galaxies are simply the oldest, residing in the most old and dense haloes, meaning that the observed trend does not necessarily imply AGN-driven quenching. Further confirmation from simulations would be needed to confirm the cause of this quenching.

Figure~\ref{quenching_trend} shows that the correlation between the fraction of quenched galaxies and \p{} decreases as the degree of smoothing increases. This suggests that the mechanisms driving the quenching of galaxies operate on smaller spatial scales. While the voxel resolution of $\rho_0$ is $0.7~\mathrm{Mpc} / h$, which roughly corresponds to the scale of a cluster-sized halo, the \texttt{BORG} initial conditions are constrained with a voxel resolution four times larger. This could imply that the quenching of low mass galaxies, with their strong positive correlation at $\rho_0$, is directly related to the internal processes on the scale of galaxy clusters. This is consistent with the correlations discussed in Section~\ref{sec:sfr_discussion}, and evident for the NSA results in Figure~\ref{fig:spearman_correlations_comparison}: galaxies in higher density environments have lower \HI{} mass content and star formation rates, likely due to the processes such as gas heating, strangulation, and the effects of galaxy group environments. We note that the analogous correlations are significantly weaker for the ALFALFA sample, which is dominated by gas-rich, star forming galaxies and therefore spans a much narrower range in SFR and colour. The strongest correlation between \p{} and quenched fraction is $\rho_0$ supporting the idea that the internal processes of a galaxy cluster play a very important role in the quenching of a galaxy.

Figure~\ref{fig:quenching_comparison} shows similar trends between the quenched fraction of galaxies in bins of $\rho_2$ and in bins of \df. Both show weaker correlations between environment and high mass galaxies, as expected, and both show steep correlations between low mass galaxies and environment. The most notable difference between the two, is that the relationship between low mass galaxies and \df{} begins to flatten at around 15 Mpc from the filament. In comparison, the trend with $\rho_2$ continues to increase with density. This suggests that after a certain distance from filament, the environmental processes linked to quenching are no longer effective. This can be explained by the relationship between \df{} and \p{}, which we show in Figure~\ref{fig:density_vs_distance}. There is a steep density gradient close to filaments that begins to flatten at around 15 Mpc from the filament and continues to do so up to 40 Mpc from the filament. \citet{2024MNRAS.532.4604W} find that the maximum density gradient is at 1 Mpc from the filament centre, which suggests that this is where we could expect to see the biggest impact of ram pressure stripping at least. This could suggest that \p{} is the dominant factor in the environmental quenching of galaxies, as it is the density gradient associated with a filament that is causing quenching, rather than the mechanisms associated with the filamentary structure itself.

\subsubsection{Morphology and size}

We take the S\'ersic index as analogous to the morphology of a galaxy where a S\'ersic index of $\lesssim  1.5$ approximately corresponds to disk galaxies, whereas a S\'ersic index of $\gtrsim 3$ corresponds to elliptical galaxies. Therefore, our results for NSA correlations suggest that as density increases, the fraction of elliptical galaxies increases. This is in agreement with the theory that elliptical galaxies often form in a ``hierarchical'' formation scenario through mergers between galaxies~\citep{De_Lucia_2006}. Mergers are more likely in high density environments due to the increased frequency of close galaxy encounters (see e.g.~\citealt{refId0}). The correlation of the S\'ersic index with \df{} supports this idea, suggesting that the closer a galaxy is to a filament, the more likely it is to be elliptical. However, this correlation is weaker than those with \p{}, suggesting  that \p{} is the dominant factor in influencing the morphology of a galaxy.

The correlation coefficients we find could imply that galaxies of a smaller radius are generally located in higher density areas, particularly when controlling for $M_\star$. However the statistical significance of the correlations for size is low,  rendering our results as inconclusive; there is no significant correlation between size and \p{} or \df{}. Nevertheless, this is an interesting result considering conflicting conclusions in the literature.~\citet{Ghosh__2024} investigate the relationship between galaxy size and environment at higher redshifts ($z \geq 0.3$), reporting that \textit{larger} galaxies are more likely to be found in higher density environments, independent of galaxy morphology or stellar mass. This is consistent with~\citet{2025arXiv250104084M}, who find that low mass galaxies ($M_\star < 10^9 M_\odot$) are larger if they are experiencing stronger tidal influences (consistent with a higher density environment). This, however, contrasts with our findings at a lower redshift.~\citet{Ghosh__2024} provide a useful analysis of the conflicting galaxy size-environment results throughout the literature. They argue that size is dependent on morphology and $M_\star$, meaning that these factors should be controlled in order to understand the correlation with environment. This could explain the low strength and statistical significance for our correlation between size and environment.

Other studies focusing on the local Universe also show conflicting results.~\citet{2013ApJ...779...29H} find that early type galaxies show no correlation between their mass-size relation and environment, while~\citet{2013MNRAS.428..109S} find that spheroids in more massive host haloes, where a spheroid is defined as an elliptical galaxy or a galaxy with a large central bulge, have a larger size at fixed $M_\star$. It is worth noting that these studies rely on observations or semi-analytical models, whereas our analysis uses constrained N-body simulations, which could account for some differences.

\subsubsection{The relative importance of \p{}, \df{}, and $\sigma$}

Throughout the results and discussion, we compare the correlations found for galaxy properties with \p{} to those found for galaxy properties with \df{}. We find consistently stronger and more statistically significant correlations with \p{} than with \df{}. Additionally, the middle row in Figure~\ref{fig:spearman_correlations_comparison}, shows that controlling for \df{}, gives stronger and more statistically significant correlations than controlling for \p{}, suggesting that the effects of \p{} are more significant than the effects of \df{}, when considered in isolation. This implies that macroscopic processes occurring due to the structure of the cosmic web itself may not be as relevant in influencing galaxy properties as the processes occurring due to \p{}. The implications of Figure~\ref{quenching_trend} are that quenching is driven more by \p{} than by \df{} (explored further in Section~\ref{sec:quenching_discussion}). These findings bolster the need for constrained simulations in studies of galactic environment, as $\rho_\text{DM}$ is otherwise much harder to estimate from the galaxy field than cosmic web structure.

Our analysis compares the effects of different values of $\sigma$, representing the degrees of Gaussian smoothing applied to \p{} (see Section~\ref{sec:density_stats}). Figure~\ref{fig:spearman_correlations_comparison} shows that most correlations peak in strength and statistical significance with smoothing scales below 4~Mpc/$h$. This provides insight into the scale of interactions that are important in contributing to a galaxy's properties, and is consistent with other studies such as~\citet{Lovell_2021} and~\citet{wu2024galaxyhalo}. We can compare these results to the findings from \citet{2024MNRAS.529.3446C}, who suggest that a galaxy’s properties are primarily dictated by its age and star formation history, with environment playing a secondary role. However, since halo density influences the timing of galaxy formation, the impact of environment is still interconnected with early formation processes, rather than being a purely external factor acting later in a galaxy’s evolution.

\subsection{Comparison to Literature}\label{sec:litrev}

This section reviews methods for quantifying environment in galaxy studies which focus on $N$-body simulations, hydrodynamical simulations, and observational survey data.

\subsubsection{Prior work with constrained N-body simulations}

Constrained $N$-body simulations such as \texttt{ELUCID}~\citep{Wang_2016} and \texttt{CSiBORG}~\citep{10.1093/mnras/stac2407,Bartlett_2022,Stiskalek_2024}, provide a novel framework to study the galaxy and halo relationship with environment. \citet{Wang_2018} investigate the relationship between galaxy quenching and matter density, finding that quenching correlates with $M_\star$, halo mass, and large-scale matter density, which is consistent with our results.
\citet{Zhang_2021} find that galaxies with a characteristic subhalo mass below $\sim 10^{12}M_{\odot} h^{-1}$ exhibit a redder colour in nodes, while above the characteristic mass there is no environmental dependence. Our findings align with these results, as we find a negative correlation between colour and \p{}. We also find that at high $M_\star$, galaxy quenching has a lower correlation with environment than at low $M_\star$, suggesting that internal galaxy processes dominate environmental effects at this scale.

\citet{xu2023connectionsdssgalaxieselucid} report a correlation of 0.14 between the colour ($g - r$) of SDSS galaxies, and $\delta_{2,1}$, the matter density contrast for subhaloes in the \texttt{ELUCID} simulation, at a Gaussian smoothing of $2.1~\mathrm{Mpc} / h$. Comparatively, we find a correlation of 0.12 between colour ($u - r$) and $\rho_2$. Although the measures of environment and colour differ slightly, these correlations show a good degree of consistency. It should however be noted, that the former value does not include an estimation of uncertainty.

An advantage of \texttt{CSiBORG} in comparison to other constrained simulations such as \texttt{ELUCID}, is the availability of 101 realisations, which quantify the reconstruction uncertainty. Additionally, with \texttt{CSiBORG} we compare our results to unconstrained simulations. This allows us to quantify deviations from the null hypothesis that there would be no correlation between a random density distribution and galaxy properties. From this, we produce the significance values in Figure~\ref{fig:spearman_correlations_comparison}.

\subsubsection{Other observational studies}

There is substantial literature investigating the correlation between galaxy properties and environment by using observational evidence only (see e.g.~\citealt{Douglass_2017,2018MNRAS.474..547K,hoosain2024effectcosmicwebfilaments,tudorache2024mighteehistarformingpropertieshi,vankempen2024comprehensiveinvestigationenvironmentalinfluences}). Such studies often quantify environment by using algorithms such as \texttt{DisPerSE} to map the cosmic web structure based on observed (redshift space) galaxy distribution, or by directly estimating the matter density from observational data. For example, both~\citet{2018MNRAS.474..547K} and~\citet{hoosain2024effectcosmicwebfilaments} use \texttt{DisPerSE} to map the cosmic web, comparing the distance of a galaxy to cosmic web structures with a galaxy's properties.~\citet{2018MNRAS.474..547K} find that more massive and redder galaxies tend to lie closer to filaments and sheets, and that redder galaxies are also preferentially found nearer to nodes. Importantly, they show that even within the star-forming population, galaxies become redder (or have lower specific star formation rates) as they approach filaments, indicating that some quenching occurs in these environments, independent of stellar mass. They compare their observational findings to the \texttt{Horizon}-AGN simulation~\citep{HAGN}, finding qualitative agreement. Similarly,~\citet{hoosain2024effectcosmicwebfilaments} find that more massive galaxies tend to lie closer to filaments, and that, independent of stellar mass, galaxies near filaments are redder and more gas deprived. They attribute these trends to the increasing prevalence of galaxy group environments around filaments. Furthermore, for galaxies in low mass haloes, they find that filaments have an additional effect on their gas content. Figure~\ref{fig:spearman_correlations_comparison} highlights the agreement of our results with these findings. In particular we see redder galaxies and lower $M_{\HI}/M_\star$ content closer to filaments (or in higher density environments). In addition,~\citet{tudorache2024mighteehistarformingpropertieshi} analyse the star formation histories from observational data, searching for a link between \HI{} properties and star formation. While they find no significant correlation between peak star formation activity and filament proximity, they note that the two galaxies in their sample located within 1 Mpc of a filament spine are \HI{} deficient and have a very low gas-depletion timescale, suggesting a potential link between a galaxy's properties and its distance to filament. 

\subsubsection{Comparison to simulations}

Correlations between galaxy properties and environment are often studied in hydrodynamical simulations.
\citet{Donnan_2022} combine observational analyses with \texttt{IllustrisTNG} simulations to find that galaxies closer to nodes have a $\sim$0.02 dex higher gas-phase metallicity at fixed stellar mass than those further away, to a $5\sigma$ significance level with Spearman correlation statistics. In comparison, we find a very low correlation between metallicity and environment, likely because~\citet{Donnan_2022} have used a mass-metallicity measurement that is only available for star-forming galaxies, whereas we considered metallicity across the whole NSA database (with metallicity data that should be ``used with caution,'' see Section~\ref{sec:NSA_data}).

\citet{ma2024neutraluniversemachinefilamentsdarkmatter} use \texttt{IllustrisTNG} together with NeutralUniverseMachine to study how distance to filament affects cold gas content. They find that the role of filaments in affecting \HI{} content is generally much weaker than the effect of halo environment. This is consistent with our result that \p{} is more significant in affecting galaxy properties than \df{}.

Hydrodynamical simulations have also been used to investigate the star-forming properties of galaxies (see e.g.~\citealt{quenching_paper, Ko_2024, byrohl2024introducingcosmostngsimulatinggalaxy}).
For example,~\citet{quenching_paper} find lower mass galaxies to be more susceptible to environmental quenching and~\citet{Ko_2024} look at how large-scale structures influence star formation in galaxy clusters.~\citet{Ko_2024} work in the redshift range $0.3 < z < 1.4$, and find that clusters well-connected to surrounding filaments and groups tend to have a lower quenched fraction of galaxies. This contrasts our result that quenched fraction increases in denser environments, and is likely due to the different redshift regimes: at high redshift filaments can feed clusters, while at low redshift high density environments tend to cause quenching.

Hydrodynamical simulations have been useful for studying the scale at which environmental properties are most important. This question is directly analogous to our interest in the most statistically significant degree of Gaussian smoothing. Figure 1 shows visually how Gaussian smoothing is applied in the CSiBORG simulations. \citet{wu2024galaxyhalo} use \texttt{IllustrisTNG} to find the connection between galaxies, dark matter haloes and their large-scale environment. They use the \texttt{DisPerSE} algorithm to quantify the distance from cosmic web features, and conclude that overdensity on $2~\mathrm{Mpc} / h$ scales is the most significant factor (see their Figure 3). The statistical significance of our results also generally peaks at around 2~Mpc/$h$, in good agreement with~\citet{wu2024galaxyhalo}. Similarly,~\citet{Lovell_2021} report that overdensity on a 2-4~Mpc scale is more important than 1 or 8~Mpc scales, further supporting this conclusion. 

\citet{storck2024causaleffectcosmicfilaments} probe the linear and non-linear effects of large-scale environment on dark matter halo formation. They find that the mass and virial radius of Milky Way-mass haloes are not significantly affected by environment, with sub-percent variations. By contrast, the orientation of the halo and its angular momentum relative to the nearest filament is highly sensitive to environment, with changes of $10-80\%$, and these orientation trends depend on distance to the filament.

Constrained $N$-body simulations are the way forward in studying galaxy formation and evolution. They offer a comprehensive comparison of the factors affecting the local Universe, whereas no study performed within the scope of a hydrodynamical simulation is guaranteed to represent the observations due to the inherent challenges of galaxy formation modelling. Observational data is essential for validating these models, but constrained simulations overcome the limitations posed by incompleteness and uncertainty quantification that arise when defining the cosmic environment directly from the observed galaxy distribution. 

\subsection{Future work}

There are some known limitations of the \texttt{CSiBORG} suite of simulations introduced in~\cite{Bartlett_2022} based on the initial conditions of~\cite{Jasche_2019}. This version of \texttt{CSiBORG} is known to overpredict cluster masses and the high-mass end of the halo mass function~\citep{10.1093/mnras/stac2407} and only later iterations of \texttt{CSiBORG} have addressed this issue~\citep{2024MNRAS.527.1244S}. In this work, we assessed the systematic uncertainty of \texttt{CSiBORG} by re-running our analysis using updated, in-development versions of the \texttt{BORG} initial conditions. Following this comparison, we estimate that the systematic uncertainty introduced by \texttt{CSiBORG} is no greater than $3\sigma$. The largest of these uncertainties affects SFR and sSFR, but their correlations remain well above the $5\sigma$ significance level. Future applications of \texttt{CSiBORG} to study the galaxy-environment relation would benefit from improved \texttt{BORG} initial conditions and self-consistently deriving the real-space distance using the~\texttt{CSiBORG} velocity field.

Research on galaxy formation and evolution could benefit from further categorising the galaxies and haloes we are studying, expanding on the ``high mass'' and ``low mass'' galaxies we evaluate in this paper. For example, a useful extension would be to categorise galaxies into \textit{satellite galaxies} and \textit{central galaxies} (see e.g.~\citealt{Wang_2018,quenching_paper}).~\citet{quenching_paper}, amongst other studies, show that central and satellite galaxies often show clear differences in their physical properties. They report that central galaxies and high mass satellite galaxies experience quenching due to internal processes while low mass satellites experience quenching due to environmental effects. ~\citet{Wang_2018} also find differences between the quenching of satellite and central galaxies, reporting that satellite galaxies show a strong correlation with environmental density, whereas central galaxies show little to no correlation. When combining these results, they find an overall correlation between environmental density and the fraction of quenched galaxies, which is in agreement with the correlations found in this paper. However, their analysis highlights that there could be nuances within our own data and results which could be explored in future work. It could also be useful to categorise galaxies based on their proximity to other galaxies, differentiating between those galaxies that are members of a group and those that are more isolated (see e.g.~\citealt{vankempen2024comprehensiveinvestigationenvironmentalinfluences,hoosain2024effectcosmicwebfilaments}). This could provide further constraints on the mechanisms responsible for quenching. As discussed in Section~\ref{sec:sfr_discussion}, there is extensive literature on the causes of quenching, often with contradicting conclusions. Some state that quenching is due to the location of a galaxy in a group environment, while others argue that it is the lack of access to gas from filament arms or the matter density of a galaxy's environment. Using \texttt{CSiBORG} to investigate separately the effect of environmental density and distance to filament, while differentiating betweens galaxies that are centrals, satellites, members of a group, or in isolation, would provide useful insight into the significance of each factor.

Further to this, some studies investigate the importance of the mass of the host dark matter haloes in dictating galaxy properties, especially galaxy quenching~\citep{Wang_2018,ma2024neutraluniversemachinefilamentsdarkmatter}. Therefore, a useful application of \texttt{CSiBORG} would be to assign halo masses to galaxies and compare the effect of the halo mass on galaxy properties.~\citet{Stiskalek_2024} study the haloes derived from \texttt{CSiBORG}, showing that a halo mass of at least $10^{14} M_\odot$ is required for it to be consistently reconstructed. Thus, the highest mass galaxies can likely by unambiguously matched directly to \texttt{CSiBORG} haloes, whereas a probabilistic matching scheme would be required for lower-mass haloes.

Incorporating those analyses suggested above would require additional modelling and data processing. In particular, identification of centrals and satellites, as well as halo mass assignment, would require either probabilistic matching techniques or additional catalogues not currently used here. A useful first step would be to focus on the most massive haloes and highest-density regions, which can be reconstructed most reliably. We therefore reserve this for future work in order to maintain a clear scope for the present analysis.

It would also be useful to apply our method to hydrodynamical simulations, in order to compare their galaxy--environment correlations to observational results. Since the dark matter density field is easily obtainable in simulations, our method of defining environment provides a better comparison to simulations than those based on observed galaxy catalogues. In addition, applying our method to a hydrodynamical simulation would allow us to investigate the distribution we find in $d_{\rm fil}$. This is important because generally hydrodynamical simulations don't find  $d_{\rm fil}$ for galaxies to be greater than 15 Mpc (e.g. \citealt{2018MNRAS.474..547K}), whereas many observational studies find much larger separations (e.g. \citealt{2017A&A...599A..31H,hoosain2024effect,2025MNRAS.539.2362J}). Investigating this difference could help clarify whether our $d_{\rm fil}$ values agree with expectations from simulations, or whether they are affected by factors such as differences in tracer number density.

Lastly, upcoming surveys such as the Large Synoptic Survey Telescope (LSST;~\citealt{2019ApJ...873..111I}), Euclid~\citep{2011arXiv1110.3193L}, and the Square Kilometre Array (SKA;~\citealt{2015arXiv150104076M}) will increase galaxy sample sizes by orders of magnitude and provide unprecedented data quality. Moreover, these datasets will improve sampling of low mass galaxies, which are under-represented in current surveys due to their faintness. This will enable a more complete analysis of the relationship between galaxy properties and their environment.

\section{Conclusion}\label{sec:conclusion}

\begin{table}
  \centering
  \begin{tabularx}{\linewidth}{lXcc}
    \toprule
    Spearman Correlation & Survey &  Coefficient & Significance\\
    \toprule
    $C(\text{colour},\rho_4|M_\star)$ & NSA & 0.07 & $11\sigma$ \\
    \midrule
    $C(\mathrm{sSFR},\rho_0)$ & NSA  & -0.08 & $10\sigma$ \\
    \midrule
     $C(\text{S\'ersic index},\rho_4|M_\star)$ & NSA & 0.06 & $9\sigma$ \\
    \midrule
     $C(\text{colour},\rho_0)$ & NSA & 0.07 & $8\sigma$ \\
    \midrule
     $C(\rm{colour},d_{\rm{fil}}|M_\star)$ & NSA & -0.05  & $8\sigma$ \\
    \midrule
     $C(\mathrm{sSFR},\rho_2|d_{\rm{fil}})$ & NSA  & -0.09 & $7\sigma$ \\
    \midrule
     $C(\mathrm{sSFR},\rho_2|M_\star)$ & NSA & -0.07 & $7\sigma$ \\
    \midrule
    $C(\text{S\'ersic index},\rho_{2})$ & NSA & 0.06  & $7\sigma$ \\
    \midrule
     $C({\mathrm{sSFR}},d_{\rm{fil}})$& NSA & 0.10 & $6\sigma$ \\
    \midrule
    $C({\mathrm{sSFR}},d_{\rm{fil}}|\rho_0)$& NSA & 0.09 & $6\sigma$ \\
    \midrule
     $C({\mathrm{sSFR}},d_{\rm{fil}}|M_\star)$& NSA & 0.05 & $6\sigma$ \\
    \midrule
    $C(\text{S\'ersic index}, \rho_2|d_{\rm{fil}})$ & NSA  & 0.05 & $6\sigma$ \\
    \midrule
     $C(\text{S\'ersic index}, d_{\rm{fil}}|M_\star)$& NSA & -0.04 & $6\sigma$ \\
    \midrule
    $C(\mathrm{SFR},\rho_2)$ & NSA  & -0.12 & $5\sigma$ \\
    \midrule
    $C(\text{colour},d_{\rm{fil}})$& NSA & -0.09 & $5\sigma$ \\
    \midrule
    $C(\text{colour},d_{\rm{fil}}|\rho_0)$& NSA & -0.08 & $5\sigma$ \\
    \midrule
     $C(\mathrm{SFR},\rho_2|M_\star)$ & NSA & -0.07 & $5\sigma$ \\
    \midrule
     $C(\text{S\'ersic index},d_{\rm{fil}})$ & NSA & -0.05 & $5\sigma$ \\
    \midrule
     $C(\text{Size},\rho_2|M_\star)$& NSA & -0.05 & $5\sigma$ \\
    \midrule
     $C(\text{Size},d_{\rm{fil}}|M_\star)$& NSA & 0.04 & $5\sigma$ \\
    \midrule
    $C(\text{S\'ersic index}, d_{\rm{fil}}|\rho_0)$& NSA & -0.04 & $5\sigma$ \\
    \midrule
     $C(M_{\HI{}}/M_\star,\rho_4)$ & ALFALFA & -0.10 & $8\sigma$ \\
     \midrule
     $C(M_{\HI{}}/M_\star,d_{\rm{fil}})$ & ALFALFA & 0.07 & $5\sigma$ \\
    \bottomrule
    \end{tabularx}
    \caption{Summary of Spearman correlation coefficients with a statistical significance $\geq 5\sigma$.}
    \label{tab:results_summary}
\end{table}

In this work, we quantify correlations between properties of galaxies and both the local matter density smoothed on various scales and the distance from a cosmic web filament. Galaxy properties are obtained via the NSA and ALFALFA surveys. We define environment using 101 constrained $N$-body simulations from the \texttt{CSiBORG} suite. We look specifically at correlations with colour, SFR, sSFR, stellar mass, size, S\'ersic index, metallicity and $M_{\HI}/M_\star$. Finally, we quantify the relationship between the quenching of high and low mass galaxies with their environment, by finding the correlation between quenched fraction of galaxies and their environmental density as well as their distance from a cosmic web filament.

By using \texttt{CSiBORG} we are able to measure the underlying dark matter density field, which is a more fundamental probe of environment than metrics based on observed galaxy catalogues. Applying \texttt{DisPerSE} to \texttt{CSiBORG} allows us to map out the cosmic web structure of the Universe, defining the positions of nodes, sheets, filaments and voids. We thus quantify the galaxy--environment connection, defining environment in two ways: environmental density (\p{}), and distance from filament (\df{}). The correlations with the highest statistical significance are summarised in Table~\ref{tab:results_summary}. The key conclusions of the paper are as follows:

\begin{itemize}
    \item \texttt{CSiBORG} provides a robust and reliable way to quantify environment, as it provides a complete and accurate representation of the dark matter distribution in the local Universe. A major advantage of using \texttt{CSiBORG} is that its multiple realisations characterise the uncertainty with which the field is reconstructed. Additionally, a comparison to unconstrained simulations can be used to find the statistical significance of the results.
    \item Defining the cosmic environment as the dark matter density environment of a galaxy, rather than its location in the cosmic web, produces stronger correlations and greater statistical significance. This suggests that environmental density has a greater impact on galaxy properties than the position of a galaxy in the cosmic web, as is perhaps not surprising given that the cosmic web simply summarises the density field. It also further highlights the importance of the constrained simulations: while a proxy to the cosmic web can be obtained from the observed distribution of galaxies, it is much more difficult to obtain the dark matter field due to the complex phenomenon of galaxy bias.
    \item We find statistically significant negative correlations between the environmental density and SFR, sSFR and \HI{}-mass -- stellar-mass ratio as well as positive correlations with colour and environmental density. Together, these correlations suggest that bluer, more \HI{} rich galaxies tend to be found in less dense areas. This is likely due to environmental processes which occur in high density environments, such as gas heating, strangulation, and the effects of galaxy group environments.
    \item The quenching of low mass galaxies ($M_\star<10^{10.5}M_{\odot}$) has a greater dependence on environmental density than the quenching of high mass galaxies ($M_\star>10^{10.5}M_{\odot}$). This is because low mass galaxies generally undergo quenching due to environmental factors, while high mass galaxies are more often quenched due to internal processes such as AGN feedback.
\end{itemize}

In summary, we find statistically significant correlations between properties of galaxies and their corresponding cosmic web environment. This has enabled us to gain insight into the processes which govern galaxy formation and evolution. The correlations calculated in this paper can be used to test theoretical models of galaxy formation, and enable us to place further constraints on the significance of astrophysical processes on galaxy formation. We demonstrate the ability of constrained simulations to pinpoint the relationships between galaxies and their environments, underlining their importance as a method with which to define the dark matter environments of the local Universe.

\section*{Data availability}

Data from the NSA, ALFALFA, and MPA-JHU surveys is publicly available at \url{http://nsatlas.org}, \url{https://egg.astro.cornell.edu/alfalfa/data/}, and \url{https://wwwmpa.mpa-garching.mpg.de/SDSS/DR7/}.  Other data underlying the article will be made available upon reasonable request.

\section*{Acknowledgements}

We thank Jens Jasche, Madalina Tudorache and Adriano Poci for useful inputs and discussion.

CG, TY and MJ acknowledge support from a UKRI Frontiers Research Grant [EP/X026639/1],
which was selected by the ERC. CG also acknowledges support from the Oxford University Astrophysics Summer Research programme. RS acknowledges financial support from STFC Grant No. ST/X508664/1, the Snell Exhibition of Balliol College, Oxford, and the CCA Pre-doctoral Program.
HD is supported by a Royal Society University Research Fellowship (grant no. 211046).

This project has received funding from the European Research Council (ERC) under the European Union's Horizon 2020 research and innovation programme (grant agreement No 693024). We also thank Jonathan Patterson for smoothly running the Glamdring Cluster hosted by the University of Oxford, where the data processing was performed.


\bibliographystyle{mnras}
\bibliography{mphysbibliography}

\end{document}